\documentclass[twocolumn,iop,revtex4,usenatbib]{openjournal}
\usepackage[T1]{fontenc}

\usepackage[english, english]{babel}

\DeclareRobustCommand{\VAN}[3]{#2}
\let\VANthebibliography\thebibliography
\def\thebibliography{\DeclareRobustCommand{\VAN}[3]{##3}\VANthebibliography}

\usepackage{algorithm} 
\usepackage[noend]{algpseudocode} 

\usepackage[bitstream-charter]{mathdesign}
\usepackage[usenames]{color}
\usepackage{xcolor}
\usepackage{soul}
\usepackage{multirow}
\usepackage{graphicx}
\usepackage{amsmath}
\usepackage{nicefrac}
\usepackage{microtype}
\usepackage{amssymb}
\usepackage{bbding}
\usepackage{csquotes} 
\usepackage{float}
\usepackage{xurl}
\usepackage{adjustbox}
\usepackage{comment}
\usepackage[hyperfootnotes=false]{hyperref}
\usepackage{tensor} 
\usepackage{enumerate}
\usepackage{orcidlink} 

\hypersetup{colorlinks=true, citecolor=blue, linkcolor=blue, urlcolor=blue}
\definecolor{hotpink}{RGB}{255, 105, 180}

\definecolor{orcidlogocol}{HTML}{A6CE39}

\DeclareSymbolFont{usualmathcal}{OMS}{cmsy}{m}{n}
\DeclareSymbolFontAlphabet{\mathcal}{usualmathcal}

\definecolor{teal}{RGB}{0,128,128}

\graphicspath{src/macro_sampler/plots}

\newcommand{\ltsima}{$\; \buildrel < \over \sim \;$}
\newcommand{\lsim}{\lower.5ex\hbox{\ltsima}}
\newcommand{\gtsima}{$\; \buildrel > \over \sim \;$}
\newcommand{\gsim}{\lower.5ex\hbox{\gtsima}}

\newcommand{\dd}{\mathrm{d}}

\newcommand{\ci}{\mathrm{i}}

\algdef{SE}[REPEAT]{Repeat}{EndRepeat}[1]{\textbf{repeat} #1 \textbf{times}}{}

\begin{document}

\title{Visualisation of spherical harmonics in Peirce's quincuncial projection\vspace{-30pt}}

\author{Bj{\"o}rn Malte Sch{\"a}fer\,\orcidlink{0000-0002-9453-5772}$^{1,2\sharp}$}
\thanks{$^\sharp$ \href{mailto:bjoern.malte.schaefer@uni-heidelberg.de}{bjoern.malte.schaefer@uni-heidelberg.de}}

\affiliation{$^{1}$ Zentrum f{\"u}r Astronomie der Universit{\"a}t Heidelberg, Astronomisches Rechen-Institut, Philosophenweg 12, 69120 Heidelberg, Germany}
\affiliation{$^{2}$ Interdisziplin{\"a}res Zentrum f{\"u}r wissenschaftliches Rechnen, Universit{\"a}t Heidelberg, INF205, 69120 Heidelberg, Germany}

\begin{abstract}
The spherical harmonics $Y_{\ell m}(\theta,\varphi)$ are complex-valued functions on the surface of a sphere, and have found widespread application in physics and astronomy. Every physics students knows them from quantum mechanics and electromagnetic theory, where they form the basis of hydrogen orbitals and of the multipole expansion, respectively. More advanced applications include the physics of the cosmic microwave background, gravitational lensing, and gravitational waves. In this paper I aim to contrast their usual $3d$ visualisation with Peirce's quincuncial projection, a conformal projection of the sphere onto a $2d$ unfolded square dihedron, where the projection respects the fundamental rotational symmetries and preserves angles. With this mapping, I guide the reader through the properties of the spherical harmonics in a pedagogical way and show that many of their mathematical relations have an intuitive visualisation on Peirce's $2d$ map, which might be useful for people challenged by processing $3d$ shapes, or which people might appreciate aesthetically.
\end{abstract}


\maketitle

\section{Introduction: Spherical harmonics $Y_{\ell m}(\theta,\varphi)$}

\subsection{A harmonic system on the sphere}
The spherical harmonics solve the differential equation
\begin{equation}
\Delta Y_{\ell m}(\theta,\varphi) = -\ell(\ell+1)Y_{\ell m}(\theta,\varphi),
\label{eqn_helmholtz_sphere}
\end{equation}
with integer $\ell\geq 0$ and with the boundary condition $Y_{\ell m}(\theta,\varphi+2\pi) = Y_{\ell m}(\theta,\varphi)$, where $\Delta$ is the Laplace-operator on the sphere,
\begin{equation}
\Delta = 
\frac{1}{\sin\theta}\frac{\partial}{\partial\theta}\left(\sin\theta\frac{\partial}{\partial\theta}\right) + \frac{1}{\sin^2\theta}\frac{\partial^2}{\partial\varphi^2},
\end{equation}
with the azimuthal angle $\varphi$ and the polar angle $\theta$. Because eqn.~(\ref{eqn_helmholtz_sphere}) is essentially the Helmholtz-differential equation, it provides a system of waves with wave number $\ell$ (and correspondingly, wave length $\lambda = \pi/\ell$ on the surface of the sphere. A separation ansatz $Y_{\ell m}(\theta,\varphi) = T(\theta)\times P(\varphi)$ leads to the explicit solution
\begin{equation}
Y_{\ell m}(\theta,\varphi) = 
\sqrt{\frac{2\ell + 1}{4\pi}}
\sqrt{\frac{(\ell-m)!}{(\ell+m)!}}
P_{\ell m}(\cos\theta)\exp(\ci m\varphi)
\end{equation}
in terms of associated Legendre-polynomials $P_{\ell m}(\cos\theta)$, and a complex phase $\exp(\ci m\varphi)$. Solving eqn.~(\ref{eqn_helmholtz_sphere}) requires $\ell$ to be a positive integer, and $m$ to be integer bounded by the condition $-\ell\leq m\leq +\ell$. Intuitively, the $Y_{\ell m}(\theta,\varphi)$ are waves in the azimuthal direction $\varphi$, and the amplitude is modulated through by the polynomials $P_{\ell m}(\cos\theta)$ with the polar angle $\theta$.

The spherical harmonics $Y_{\ell m}(\theta,\varphi)$ are an orthonormal system of waves on the surface of the sphere, and allow the spectral decomposition in analogy to the Fourier-transform
\begin{equation}
\psi_{\ell m} = \int_{4\pi}\dd\Omega\:\psi(\theta,\varphi) Y_{\ell m}^*(\theta,\varphi)
\label{ylm_analysis}
\end{equation}
with the solid angle element $\dd\Omega = \sin\theta\dd\theta\dd\varphi = \dd\cos\theta\dd\varphi$ for integration. The inverse operation is given by
\begin{equation}
\psi(\theta,\varphi) = \sum_{\ell=0}^\infty\sum_{m=-\ell}^{+\ell}\psi_{\ell m}Y_{\ell m}(\theta,\varphi).
\label{ylm_synthesis}
\end{equation}

There are numerically very efficient algorithms for computing these transforms, even in the limit of high $\ell$, where the transform kernel is highly oscillatory, for instance the HEALPix-software used in data analysis of the cosmic microwave background \citep{gorski1999healpixprimer, Gorski_2005}. Generalisations of $Y_{\ell m}(\theta,\varphi)$ to non-scalar functions include the spin spherical harmonics $_{s}Y_{\ell m}(\theta,\varphi)$ \citep{goldberg_spins_1967, Huffenberger_2010, 10.1063/1.526533}, the spinor spherical harmonics for fermionic wave functions in relativistic quantum mechanics, and spin spheroidal harmonics if the condition of spherical symmetry is relaxed and functions defined on the surface of ellipsoids are considered \citep{PhysRevD.61.084004}.

\subsection{Applications in physics}
There is a plethora of applications in physics for the spherical harmonics: They appear in the multipole expansion of charge distributions $\rho$ in electrodynamics for solving potential problems of the Poisson-type,
\begin{equation}
\Delta\Phi = -4\pi\rho,
\end{equation}
in a convenient series with terms proportional to $1/r^{\ell+1}$ for the multipole contribution at order $\ell$, with $r$ being the distance of the observer to the charge distribution. In quantum mechanics, they appear naturally when solving the Schr{\"o}dinger equation for spherically symmetric potentials $\Phi(r)$,
\begin{equation}
\left(-\frac{\hbar^2}{2m}\Delta + \Phi(r)\right)\psi = E\psi,
\end{equation}
as exemplified by the hydrogen orbitals. More advanced applications include geophysics \citep{10.1093/gji/52.2.366} and data analysis of the cosmic microwave background, whose fluctuation pattern on the celestial sphere is conveniently decomposed in terms of spherical harmonics by virtue of eqn.~(\ref{ylm_analysis}) \citep{maino_planck-lfi_1999,lewis_21cm_2007,leistedt_3d_2015}. Likewise in cosmology, full-sky analysis of the gravitational lensing effect is described using spherical harmonics, \citep{hanson_weak_2010,hirata_reconstruction_2003,beck_lensing_2018,zaldarriaga_gravitational_1998,seljak_gravitational_1996,nishizawa_cosmic_2008,okamoto_cosmic_2003}. Extensions like the spin-spherical harmonics have been constructed to deal with non-scalar fields such as polarisation of the cosmic microwave background, the cosmic gravitational wave background or the galaxy ellipticity field, \citep{fernandez-cobos_exploring_2016, kamionkowski_quest_2015,gluscevic_-rotation_2009,adamek_lensing_2015, hall_intrinsic_2014, merkel_gravitational_2011}. A comprehensive references of properties of the spherical harmonics $Y_{\ell m}(\theta,\varphi)$ is, for instance, \citet{NIST:DLMF}.

\section{Visualisation of the spherical harmonics}
Visualisation of the spherical harmonics $Y_{\ell m}(\theta,\varphi)$ needs to take care of the fact that they are complex-valued, and that the topology of their domain is the sphere, i.e. they provide a mapping $\mathbb{S}^2 \rightarrow \mathbb{C}$, or equivalently, $\mathbb{S}^2 \rightarrow \mathbb{R}^2$. $3d$-visualisations often suffer from that fact that foreground structures cover details further in the back: This is the motivation behind this article, which illustrates properties of spherical harmonics in Peirce's quincuncial projection. There, the full solid angle is mapped onto a square, such that all details of the $Y_{\ell m}(\theta,\varphi)$ are visible, irrespective of their localisation on the sphere.

One straightforward way of visualising the complex-valued spherical harmonics is computing their absolute value $|Y_{\ell m}(\theta,\varphi)|^2 = Y_{\ell m}(\theta,\varphi)Y_{\ell m}^*(\theta,\varphi)$ and to depict this value as the distance of a surface to the origin in spherical coordinates. Alternatively, one could resort to plotting real or imaginary values: Because of the oscillatory behaviour of $\exp(\ci m\varphi) = \cos(m\varphi) + \ci\sin(m\varphi)$ one directly sees the variation in the azimuthal direction determined by the choice of $m$: One encounters $m$ maxima and $m$ minima in one circumnavigation in $\varphi$ at fixed $\theta$: The resulting visualisation directly corresponds to the atomic orbitals known from theoretical chemistry. In fact, solving the Schr{\"o}dinger equation for obtaining the wave function of an electron in the Coulomb-field of an atomic nucleus would show that it comes out proportional to $Y_{\ell m}(\theta,\varphi)$ where the indices $\ell$ and $m$ acquire the interpretation of quantum numbers. They indicate the magnitude of angular momentum and its orientation, and can be measured through the magnetic moment that is associated with the wave function of an electrically charged particle.

Phase information can be conveyed by the colouring, i.e. by cycling through the spectrum of the rainbow colours as the phase evolves from $0$ to $2\pi$. In many visualisations, this information is discarded. Instead, only positive and negative values of a real-numbered representation are given. The hue of the colouring can indicate the modulus of the complex numbers, for instance dimming the shade of colour towards black could indicate small values for $\left|Y_{\ell m}(\theta,\varphi)\right|$.

\begin{figure}[htb]
    \centering
    \begin{tabular}{c}
    \includegraphics[width=0.4\textwidth]{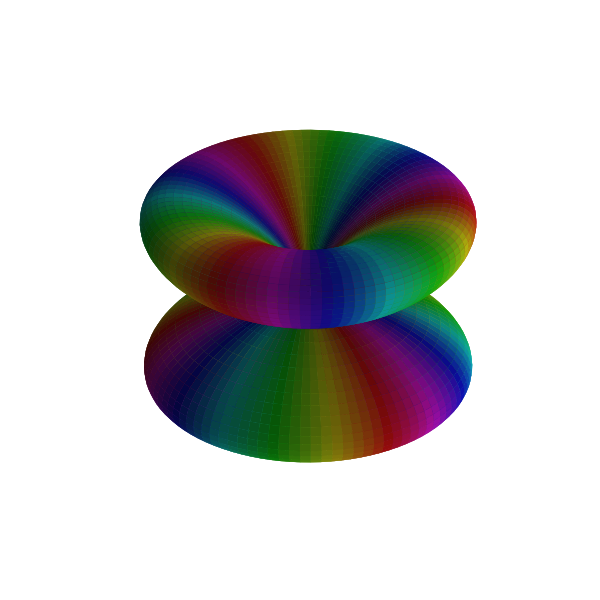} \\
    \includegraphics[width=0.4\textwidth]{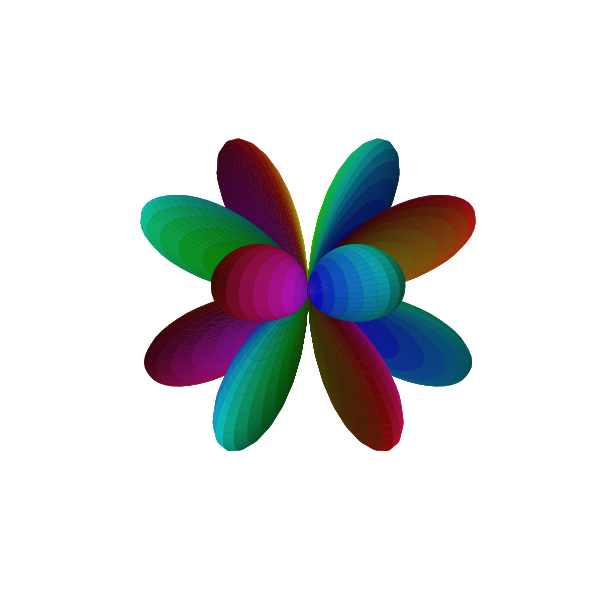} \\
    \includegraphics[width=0.4\textwidth]{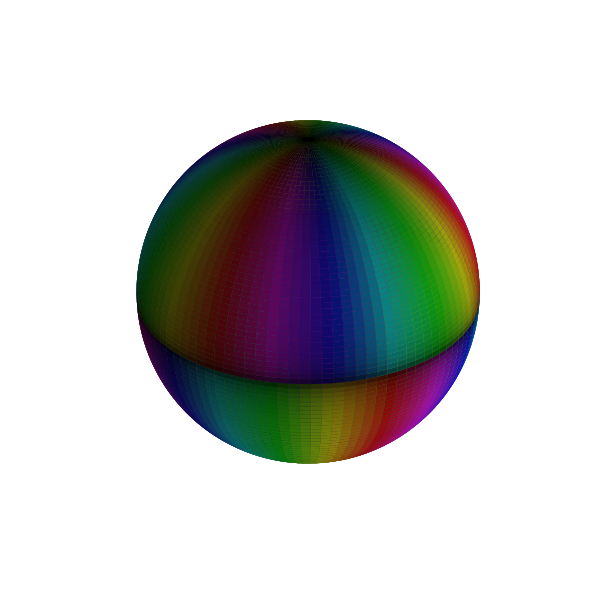}
    \end{tabular}
    \caption{Spherical harmonic $Y_{43}(\theta,\varphi)$ as an absolute value $\left|Y_{43}(\theta,\varphi)\right|$ (top), real value $\mathrm{Re}\:Y_{43}(\theta,\varphi)$ (centre) and on the unit sphere (bottom), all with the phase $\exp(\ci m\varphi)$ as colouring.}
    \label{fig:y73_3d}
\end{figure}

Common visualisations of the spherical harmonic function $Y_{43}(\theta,\varphi)$ are shown in Fig.~\ref{fig:y73_3d} in $3d$, in terms of absolute value, real value and phase. Clearly, not all information is accessible, for instance the values of the complex phase on the far side of the spherical harmonic.

\section{Peirce's quincuncial projection}
Peirce's quincuncial projection is a conformal map projection from the sphere to an unfolded square dihedron, developed by C.S.~Peirce \citep{c14fbc06-496f-3cbc-8158-f8078819cb05}. Each octant projects onto an isosceles triangle. Four of these are arranged into a diamond shape representing the Northern hemisphere, and the four remaining Southern hemisphere tiles can be added in a way to complete the diamond to a square. As such, the square forms a seamless tiling of the plane.

The projection proceeds in two steps: $(i)$ a stereographic projection from the surface of a sphere onto a tangential plane, followed by $(ii)$ a compactification onto the domain $[-1,+1]\otimes[-1,+1]$, where the latter is a ingenious application of Riemann's mapping theorem, due to \citep{Schwarz+1869+105+120} and \citep{christoffel_1867_2358602}, expertly explained by \citet{Driscoll_Trefethen_2002}. The mapping theorem states that there is a bijective, holomorphic mapping between any open, simply-connected subset of the complex plane onto the open unit disc. As a special case, Schwarz and Christoffel constructed mappings between the unit disc and regular polygons, with the particular case of a square. This mapping is conformal except at four singular, isolated points along the equator. Otherwise, in particular on the $X$-shaped region along the diagonals, the conformal factor varies only little, from which its usefulness for geography is derived: The continents on the Earth are situated in such a way that they lie roughly in the regions of low variability of the conformal factor, leading to a faithful representation of the relative size of geographic features. Regions of high variability of the conformal factor are found on the open oceans, where they would not matter visually. A funky side effect of this is that Australia and New Zealand are projected onto two different lobes and appear to be far away from each other. But when tiling the maps they again assume a sensible separation relative to each other.

\begin{figure}[htb]
    \centering
    \begin{tabular}{c}
    \includegraphics[width=0.45\textwidth]{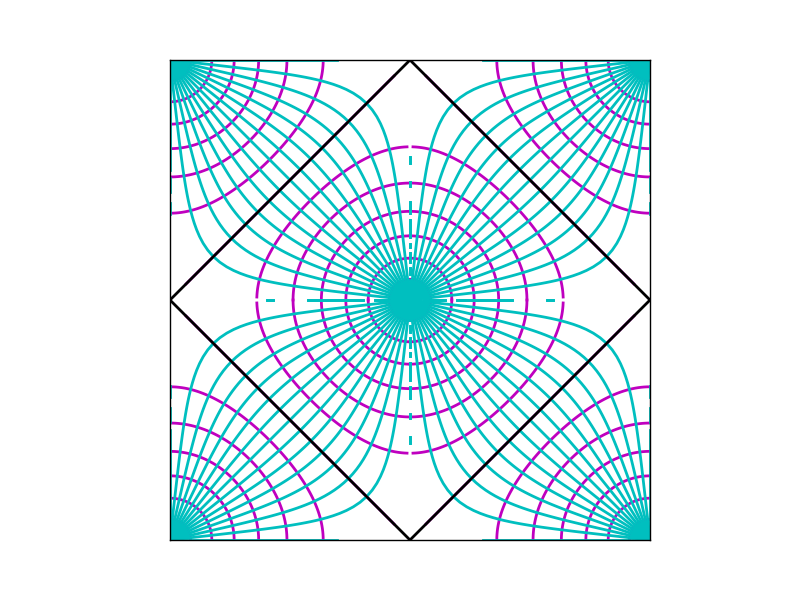} \\
    \includegraphics[width=0.45\textwidth]{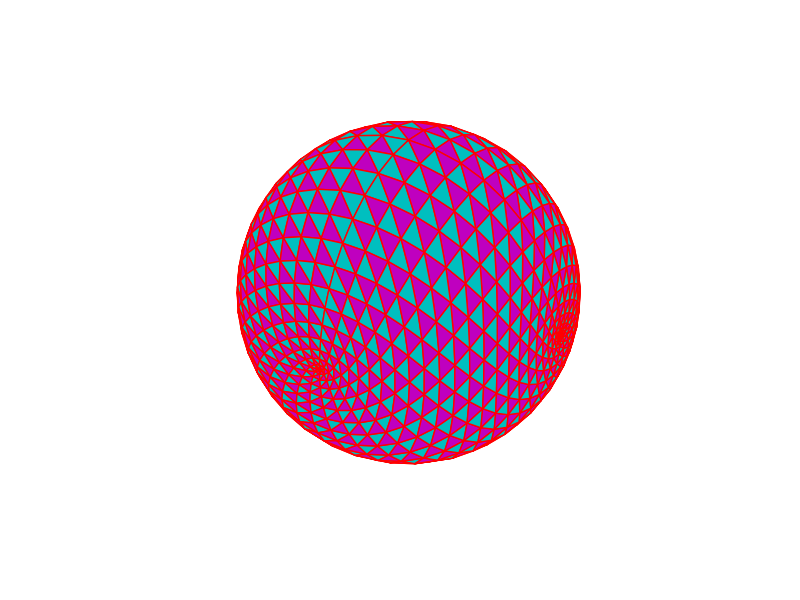}
    \end{tabular}
    \caption{The mapping of meridians and circles of latitude on a $6^\circ$-graticule in azimuth and a $15^\circ$-graticule in polar angle in Peirce's projection onto a square. The black diamond-shaped line indicates the equator (top). The inverse mapping of a Cartesian grid on Peirce's domain back onto the sphere. There are four points on the equator (two are visible in the front and two on the far side of the sphere) where the conformal factor becomes zero (bottom).}
    \label{fig:there_and_back_again}
\end{figure}

In general, the projected shapes of meridians and circles of latitude are complicated, but there are two perfectly straight meridians running from North to South pole along the diagonals of Peirce's projection, and the circles of latitude are assume more circular projected shapes close to the poles, while approximating the lozenge-shaped form of the equator at low latitude, as shown in Fig.~\ref{fig:there_and_back_again}. Due to the conformality of the mapping, they always intersect at right angles. The inverse mapping of a checker pattern in Peirce's domain back to the sphere shows that there are four points, where the conformal factor becomes zero.

\begin{figure}[htb]
    \centering
    \includegraphics[width=0.45\textwidth]{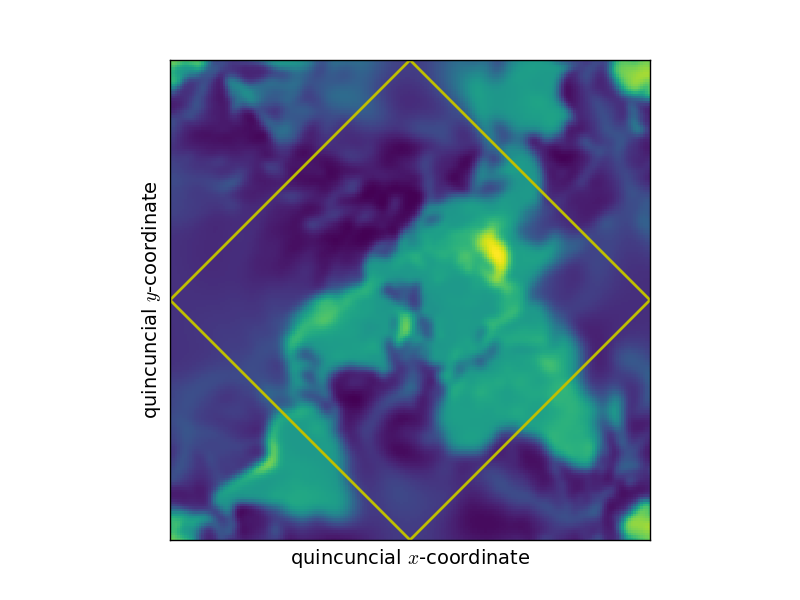}
    \caption{The Earth in Peirce's quincuncial projection, generated as a topographical map from the superposition of spherical harmonics, up to $\ell_\mathrm{max} = 100$. The yellow diamond shape is the equator in Peirce's projection. The map is smoothed with a Gaussian filter, for suppressing diffraction-like artefacts.}
    \label{fig:globe}
\end{figure}

Fig.~\ref{fig:globe} illustrates two key results at the same time, with the Earth as a well-known example: First, it shows how the entire surface gets projected onto the square according to Peirce's quincuncial projection. Second, it shows how the spherical harmonics analyse all features of the function $\psi(\theta,\varphi)$, here topographical height, as a function of scale $\lambda = \pi / \ell$ (in radians) in evaluating the transformation eqn.~(\ref{ylm_analysis}). 

When assembling $\psi(\theta,\varphi)$ from $\psi_{\ell m}$ according to eqn.~(\ref{ylm_synthesis}), information on successively smaller and smaller scales is added, such that finer and finer geographical details are visible. A rough estimate can make this more qualitative: $\ell = 100$ corresponds to features of size $\lambda = \pi/100$, i.e. $\sim 2^\circ$. With 60 arcminutes in a degree, one arcminute being the angle subtended by a nautical mile on the surface of the Earth, and the conversion of 1852~meters per nautical mile these particular features are of size $\sim 200$~km, so continents are clearly visible. Topographical details like mountain ranges or the Mediterranean Sea are resolved, but for individual islands in the Caribbean Sea or in the Mediterranean $\ell_\mathrm{max}$ would need to be larger still.

The implementation by \citet{DBLP:journals/corr/abs-1011-3189, fong2024analyticalmethodssquaringdisc} that computes spherical coordinates $(\theta,\varphi)$ from quincuncial coordinates $(x,y)$ in the domain $[-1,+1]\otimes[-1,+1]$ forms the basis of this article. Explicit conversion formulas are given in the Appendix. The inverse projection is given by \citet{solanilla_mapping}.

\subsection{Zonal, sectoral and tesseral $Y_{\ell m}(\theta,\varphi)$}
\begin{figure*}[htb]
    \centering
    \begin{tabular}{cc}
    \includegraphics[width=0.45\textwidth]{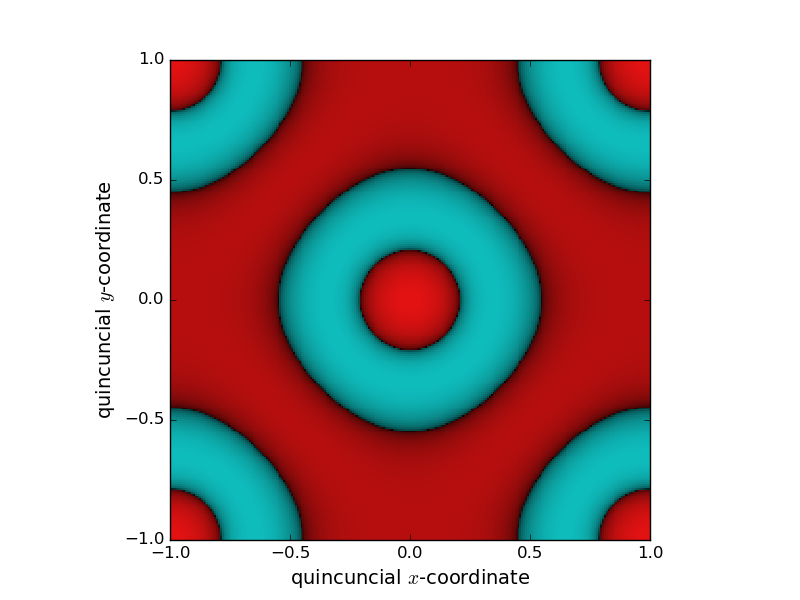} &
    \includegraphics[width=0.35\textwidth]{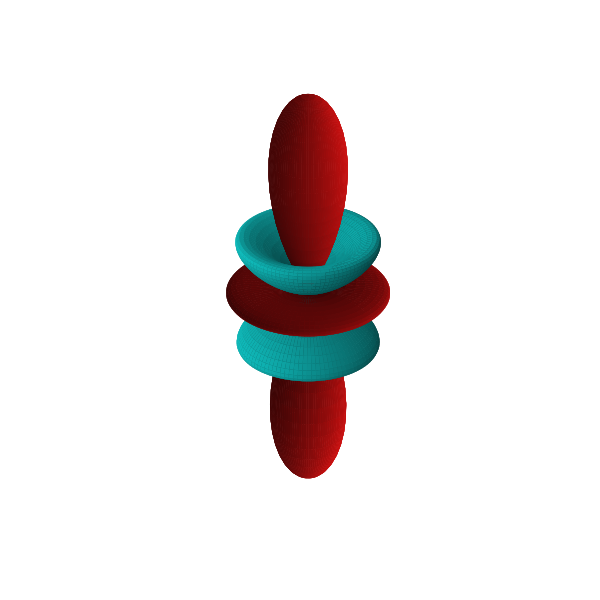}
    \end{tabular}
    \caption{Zonal spherical harmonic $Y_{40}(\theta,\varphi)$, for $\ell = 4$ and $m = 0$, in a representation with $3d$ orbitals (right) and in comparison with Peirce's projection (left).}
    \label{fig:zonal}
\end{figure*}

The spherical harmonics can be classified according to the value of $m$ for a given $\ell$, as they do change their character perceivably depending on $m$: This classification has as categories zonal ($m = 0$), sectoral ($\ell = +m$ or $\ell = -m$) and tesseral in all other cases. The zonal $Y_{\ell 0}(\theta,\varphi)$ are real valued, as the term $\exp(\ci m\varphi)$ as the only source of complex values, is inactive for $m = 0$, and they are necessarily azimuthally symmetric, because of the same reason. This behaviour is shown in Fig.~\ref{fig:zonal}, where the spherical harmonic $Y_{40}(\theta,\varphi)$ is shown in a $3d$ representation and as a $2d$ Peirce's projection. One clearly sees the lobes at the North and South pole with equal sign, the equatorial bulge with the same sign, and the two rings at intermediate Northern and Southern latitudes.

\begin{figure*}[htb]
    \centering
    \begin{tabular}{cc}
    \includegraphics[width=0.45\textwidth]{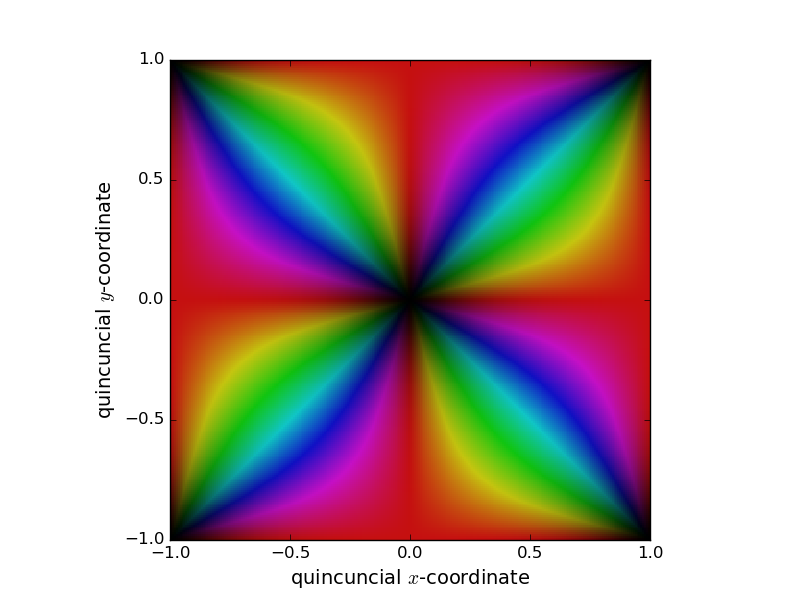} &
    \includegraphics[width=0.35\textwidth]{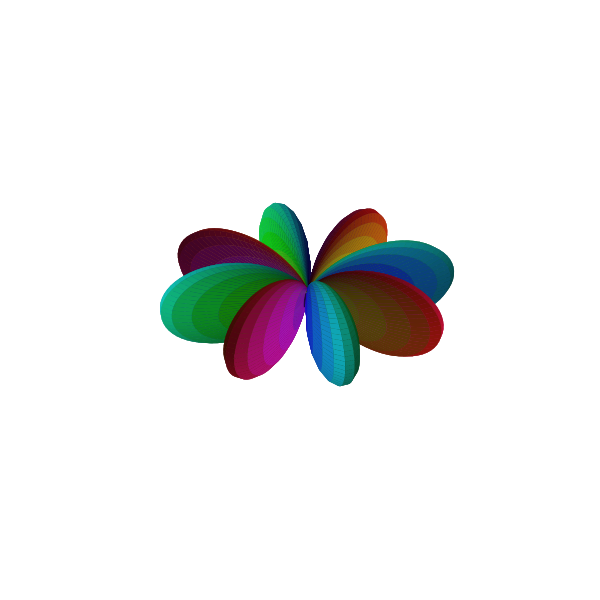}
    \end{tabular}
    \caption{Sectoral spherical harmonic $Y_{44}(\theta,\varphi)$, for $\ell = m = 4$, in a representation with $3d$ orbitals (right) and in comparison with Peirce's projection (left). In both cases, colouring indicates complex phase.}
    \label{fig:sectoral}
\end{figure*}

The sectoral spherical harmonic $Y_{44}(\theta,\varphi)$ is depicted in Fig.~\ref{fig:sectoral}: Its functional values on the North and South pole are zero, and large amplitudes appear on the equator. As $m = 4$, the complex phase cycles through the interval $0\ldots 2\pi$ exactly four times, and the real value exhibits four positive maxima and four negative minima. 

\begin{figure*}[htb]
    \centering
    \begin{tabular}{cc}
    \includegraphics[width=0.45\textwidth]{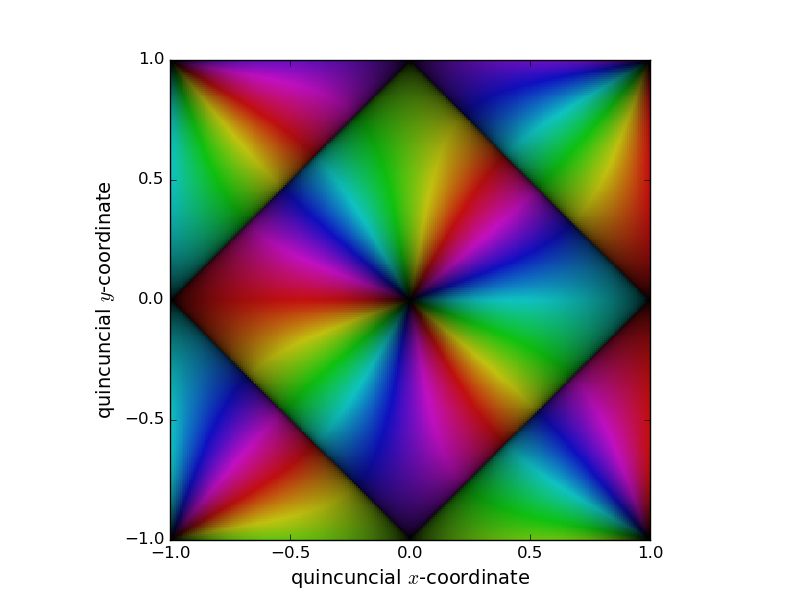} &
    \includegraphics[width=0.35\textwidth]{./y43_real.png}\\
    \includegraphics[width=0.45\textwidth]{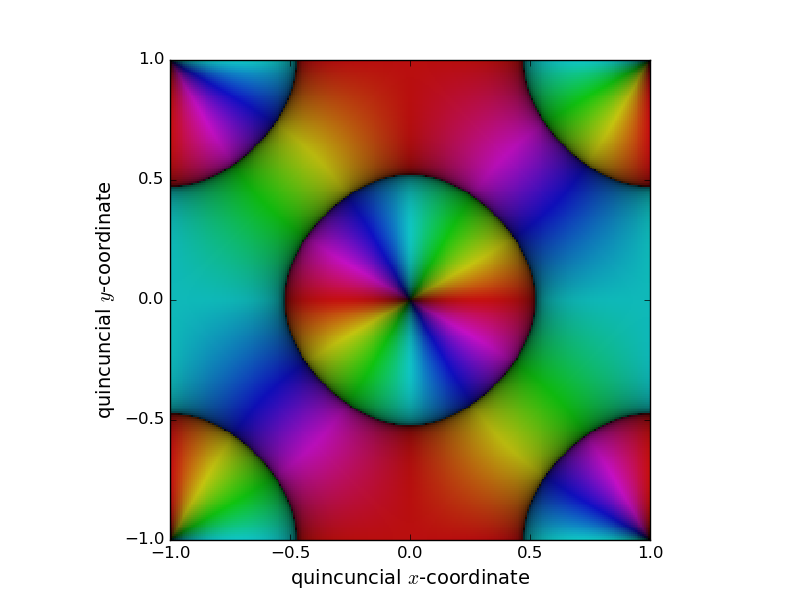} &
    \includegraphics[width=0.35\textwidth]{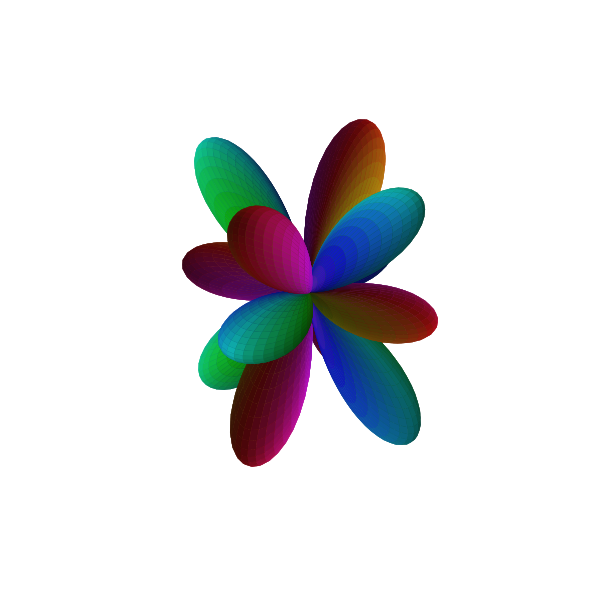}\\
    \includegraphics[width=0.45\textwidth]{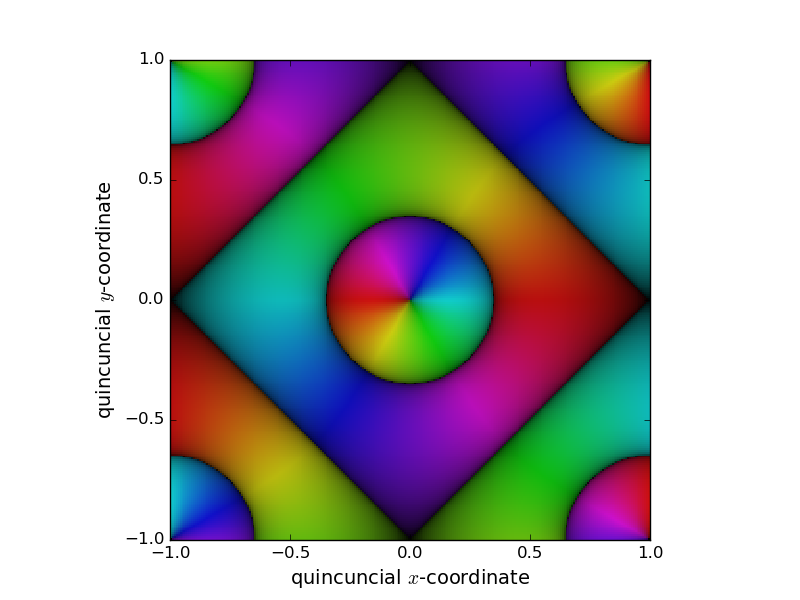} &
    \includegraphics[width=0.35\textwidth]{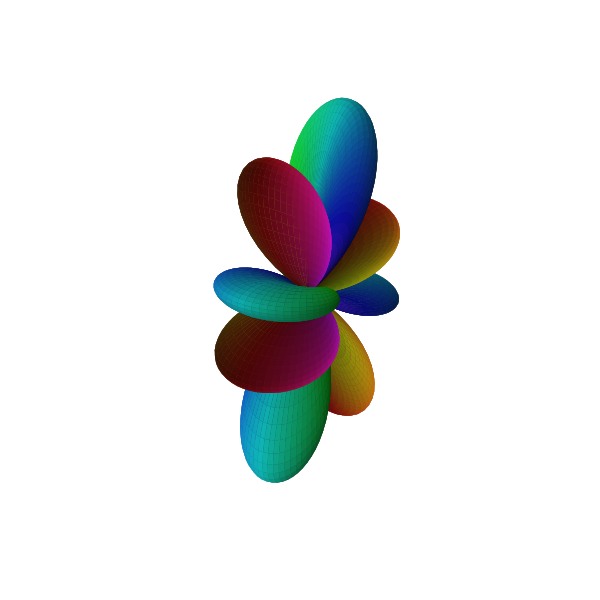}\\
    \end{tabular}
    \caption{Tesseral spherical harmonics $Y_{\ell m}(\theta,\varphi)$, for $\ell = 4$, in a representation with $3d$ orbitals (left) and in comparison with Peirce's projection (right). In both panels, colouring indicates complex phase, for $m=3$ (top row), $m=2$ (centre row) and $m=1$ (bottom row).}
    \label{fig:tesseral}
\end{figure*}

Finally, the tesseral spherical harmonics $Y_{\ell m}(\theta,\varphi)$ where $m\neq 0$ and $|m|\neq\ell$ are shown in Fig.~\ref{fig:tesseral}, specifically $Y_{41}(\theta,\varphi)$, $Y_{42}(\theta,\varphi)$ and $Y_{43}(\theta,\varphi)$. As soon as $\ell + m$ is odd, the associated Legendre-polynomials $P_{\ell m}(\cos\theta)$ assume a value of zero at the equator, which in Peirce's projection is mapped onto a diamond shape. The value of $m$ determines as well, how often the complex phase cycles through the interval $0\ldots 2\pi$: In consequence, one should see each colour $m$ times on a ring of constant latitude, and $2m$ orbitals in the $3d$ projection of the real value of $Y_{\ell m}$, $m$ with a positive sign and $m$ with a negative sign.

\begin{figure}[htb]
    \centering
    \includegraphics[width=0.45\textwidth]{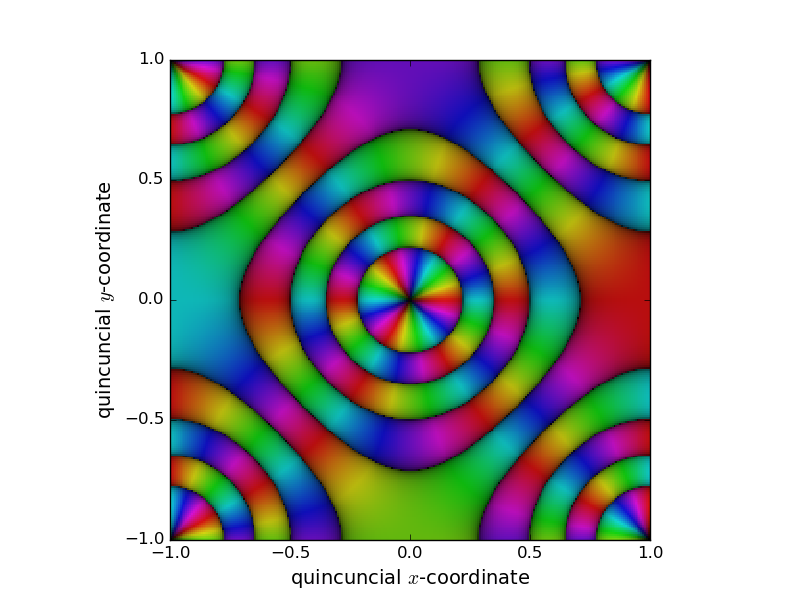}
    \caption{Spherical harmonic $Y_{11,3}(\theta,\varphi)$, with $\ell = 11$ and $m = 3$.}
    \label{fig:intermediate}
\end{figure}

Fig.~\ref{fig:intermediate} shows the spherical harmonic $Y_{\ell m}(\theta,\varphi)$ for the choice $\ell = 11$ and $m = 3$. Clearly, the associated Legendre-polynomials in $\cos\theta$ show 10 zeros between North and South pole, and each color appears three times in azimuth. The subsequent bulges of the associated Legendre-polynomials differ in sign, which is equivalent to a phase shift of $\pi$: This can be seen in the alternating colour gradients. In fact, the phase variation in azimuth is identical in every second bulge, due to $(-1)^2 = 1$.

\subsection{Visualisation of particular properties of $Y_{\ell m}(\theta,\varphi)$}

\subsubsection{Hermiticity}
\begin{figure}[htb]
    \centering
    \begin{tabular}{c}
    \includegraphics[width=0.45\textwidth]{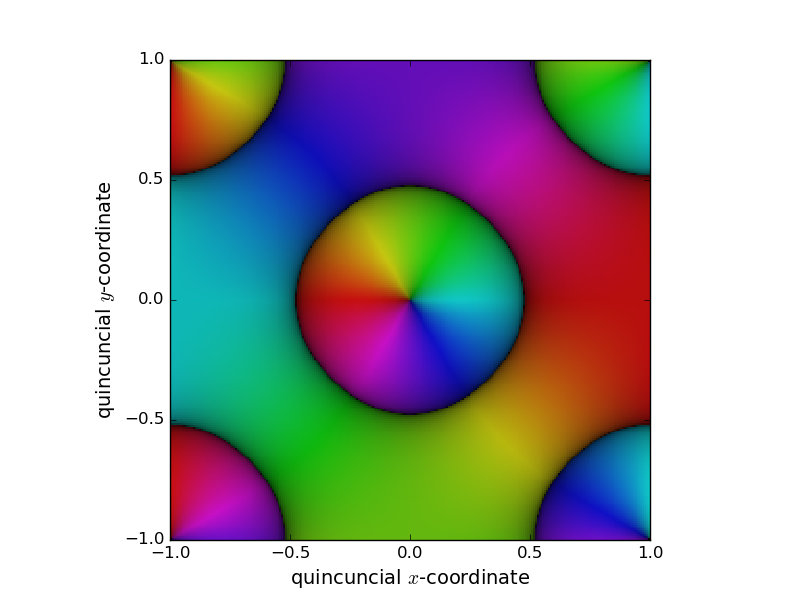} \\
    \includegraphics[width=0.45\textwidth]{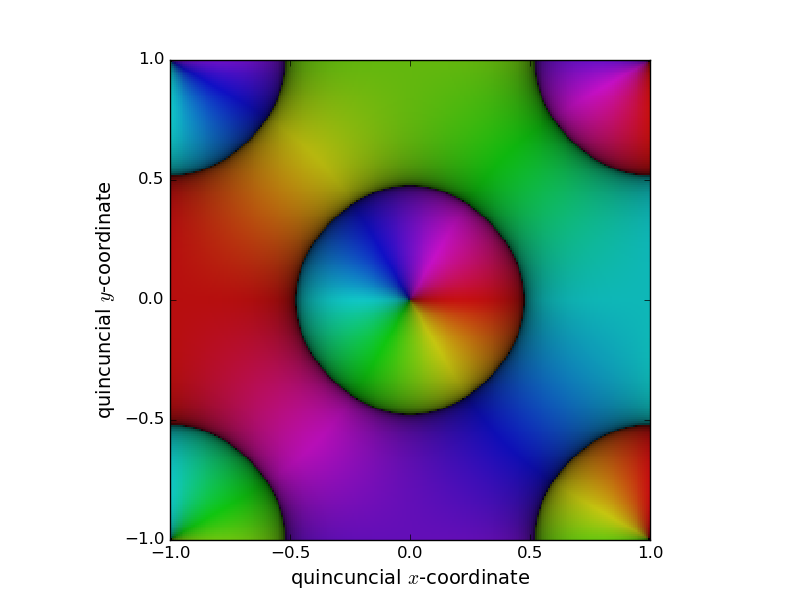}
    \end{tabular}
    \caption{Spherical harmonic $Y_{\ell m}^*(\theta,\varphi)$ (top) and in comparison $Y_{\ell,-m}(\theta,\varphi)$ (bottom), for $\ell = 3$ and $m=1$. The two plots differ by a factor $(-1)^m=-1$, illustrating the hermiticity relation.}
    \label{fig:hermiticity}
\end{figure}

Spherical harmonics as complex valued functions have simple behaviour under complex conjugation,
\begin{equation}
Y_{\ell m}^*(\theta,\varphi) = (-1)^mY_{\ell,-m}(\theta,\varphi),
\end{equation}
where one needs to change the sign of $m$ and add a pre-factor $(-1)^m$ to form the Hermiticity condition shown in Fig.~\ref{fig:hermiticity}: It compares $Y_{3,+1}^*(\theta,\varphi)$ and $Y_{3,-1}(\theta,\varphi)$ side by side, where the factor $(-1)^1 = -1$ relates the two: The places at which the colours appear are shifted in azimuthal angle $\varphi$ by $\pi$.

\subsubsection{Parity}
The values of a spherical harmonic at coordinates $\theta,\varphi$ and the antipodal point $\pi - \theta,\varphi + \pi$ are related by a prefactor of $(-1)^\ell$,
\begin{equation}
Y_{\ell m}(\pi - \theta,\varphi + \pi) = (-1)^\ell Y_{\ell m}(\theta,\varphi).
\end{equation}

\begin{figure}[htb]
    \centering
    \begin{tabular}{c}
    \includegraphics[width=0.45\textwidth]{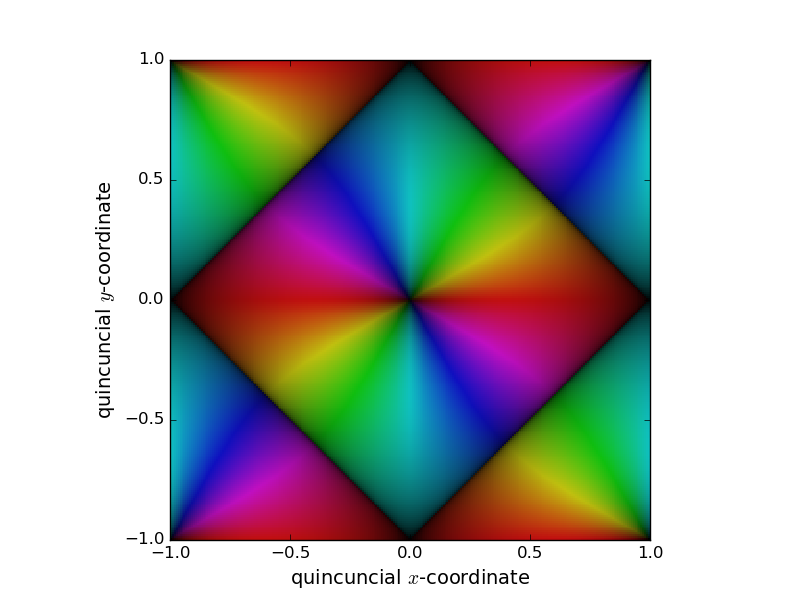} \\
    \includegraphics[width=0.45\textwidth]{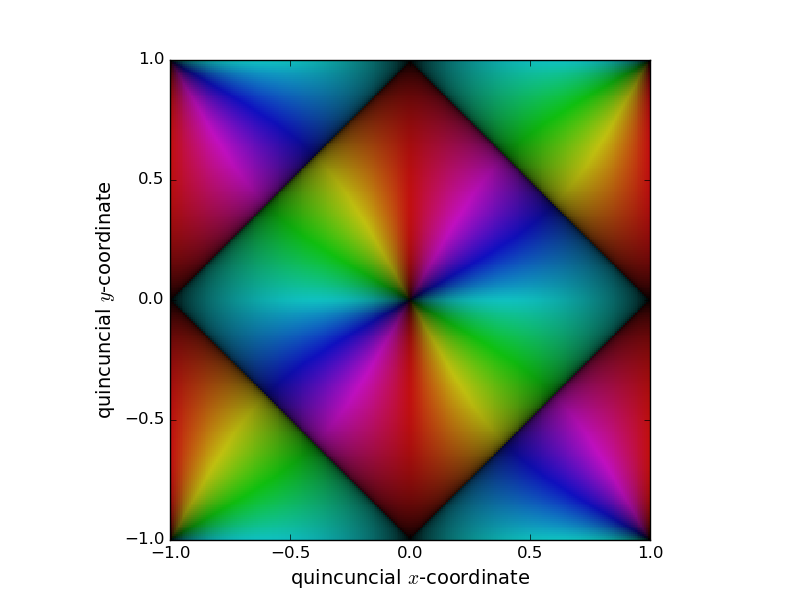}
    \end{tabular}
    \caption{Spherical harmonic $Y_{\ell m}(\pi - \theta,\varphi + \pi)$ (top) and in comparison $Y_{\ell m}(\theta,\varphi)$ (bottom), for $\ell = 3$ and $m=2$. The two plots differ by a factor $(-1)^\ell = -1$, showing the influence of parity inversion.}
    \label{fig:parity}
\end{figure}

This behaviour under parity transform is illustrated in Fig.~\ref{fig:parity}, where $Y_{32}(\pi - \theta,\varphi + \pi)$ and $Y_{32}(\theta,\varphi)$ are shown side by side. As $\ell = 3$, the prefactor relating the two functions is $(-1)^3 = -1$. It is striking to see that this prefactor shifts the colours, which repeat twice on each ring of constant latitude, by $\pi/2$ in azimuth.

\subsubsection{Orthonormality}
\begin{figure}[htb]
    \centering
    \includegraphics[width=0.45\textwidth]{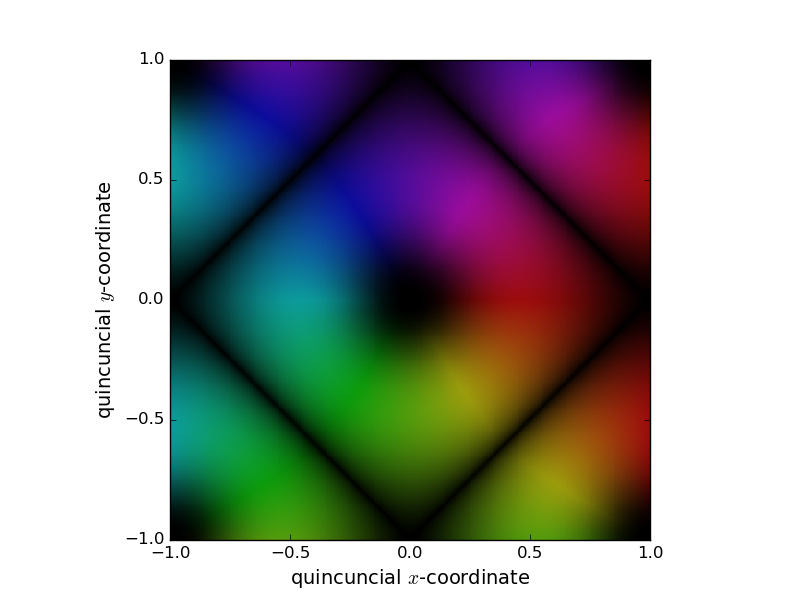}
    \caption{Product $Y_{\ell m}(\theta,\varphi)Y_{\ell^\prime m^\prime}^*(\theta,\varphi)$ for $\ell,m = 4,3$ and $\ell^\prime,m^\prime = 5,4$, for illustration of the orthonormality relation.}
    \label{fig:orthonormality}
\end{figure}

The orthonormality condition of the spherical harmonics reads
\begin{equation}
\int_{4\pi}\dd\Omega\:Y_{\ell m}(\theta,\varphi)Y_{\ell^\prime m^\prime}^*(\theta,\varphi) = 
\delta_{\ell\ell^\prime}\delta_{mm^\prime},
\end{equation}
and reflects that the integral over the solid angle $\dd\Omega = \sin\theta\dd\theta\dd\varphi$ of a pair of functions, one of which is complex conjugated, is a Hermitean scalar product. If one of the index pairs contains two unequal indices, the integral should yield zero, because the integrand shows a complete cancellation over the surface of the sphere. This is illustrated in Fig.~\ref{fig:orthonormality}, where the absolute value as a function of $\theta$ and $\varphi$ is symmetric when comparing the Northern and the Southern hemisphere, and the phase varies smoothly between $0$ and $2\pi$ in the azimuthal direction, leading to a vanishing integral.

\subsubsection{Completeness}
The completeness relation,
\begin{multline}
\delta_D(\cos\theta - \cos\theta^\prime)\delta_D(\varphi-\varphi^\prime) = \\
\sum_{\ell = 0}^\infty\sum_{m=-\ell}^{+\ell} Y_{\ell m}(\theta,\varphi)Y_{\ell m}^*(\theta^\prime,\varphi^\prime),
\end{multline}
states that there is a representation of the Dirac $\delta_D$-function in terms of the basis system. The cosines in the argument of the first term appear because the $\delta_D$-functions are normalised with respect to $\dd\Omega = \sin\theta\dd\theta\dd\varphi = \dd\cos\theta\dd\varphi$ as the integration measure.

\begin{figure}[htb]
    \centering
    \begin{tabular}{c}
    \includegraphics[width=0.45\textwidth]{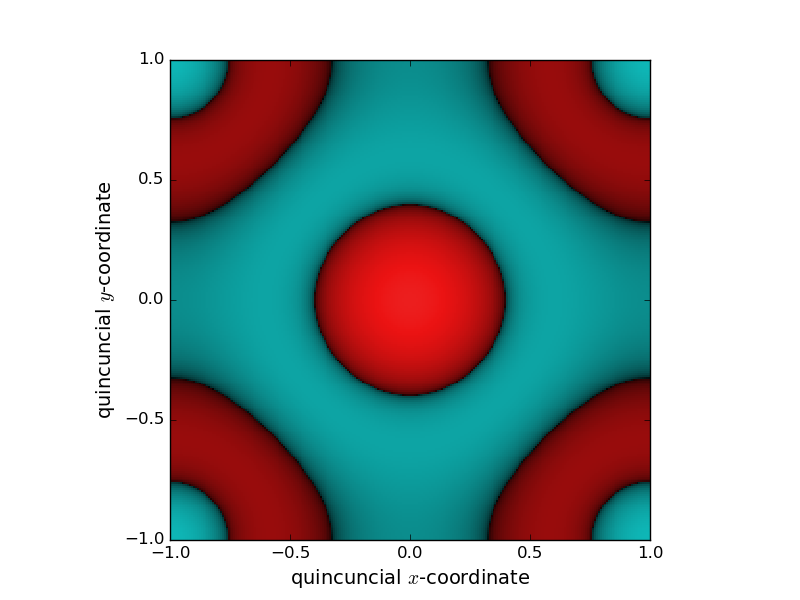} \\
    \includegraphics[width=0.45\textwidth]{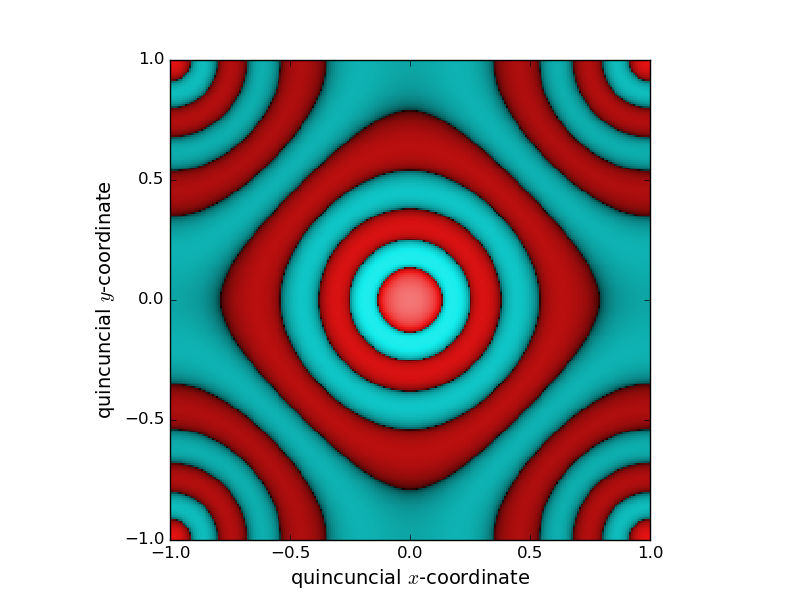} \\
    \includegraphics[width=0.45\textwidth]{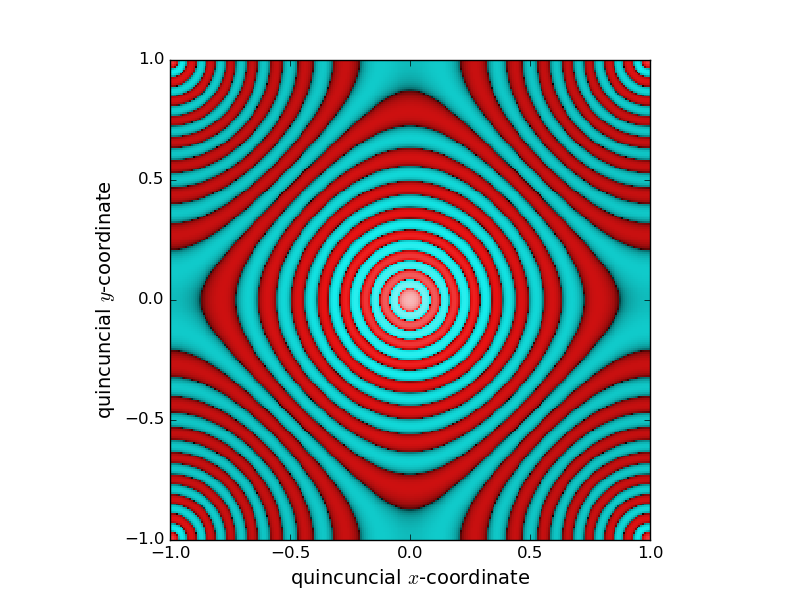}
    \end{tabular}
    \caption{Partial sums in $\ell$ of the completeness relation, for $\theta^\prime = 0 = \varphi^\prime$: In consequence, the $\delta_D$-function should appear at the North pole at the centre of the plot, successively approximated as $\ell_\mathrm{max} = 3$ (first panel), $\ell_\mathrm{max} = 10$ (second panel), $\ell_\mathrm{max} = 30$ (third panel).}
    \label{fig:completeness}
\end{figure}

For visualisation, as exemplified in Fig.~\ref{fig:completeness}, we choose $\cos\theta^\prime = 1$ and $\varphi^\prime = 0$, such that the Dirac $\delta_D$-function should appear on the North pole. Then, the completeness relation simplifies tremendously,
\begin{multline}
\delta_D(\cos\theta - 1)\delta_D(\varphi) = \\
\sum_{\ell = 0}^\infty\sqrt{\frac{2\ell+1}{4\pi}}\sum_{m=-\ell}^{+\ell} \sqrt{\frac{(\ell-m)!}{(\ell+m)!}}Y_{\ell m}(\theta,\varphi),
\end{multline}
because $Y_{\ell m}^*(\theta^\prime,\varphi^\prime) \propto P_{\ell m}(\cos\theta^\prime)\exp(\ci m\varphi^\prime) = 1$, leaving only the prefactor. As expected, there is constructive interference at the North pole, and destructive interference elsewhere, and because one always adds pairs of spherical harmonics at positive and the corresponding negative $m$, the result is real-valued.

The gradual build-up of the Dirac function at the North pole is shown in Fig.~\ref{fig:completeness} for partial sums in $\ell_\mathrm{max}$. At the same time, one sees an increasing number of rings as $\ell_\mathrm{max}$ becomes larger, due to the more violent oscillatory behaviour of the Legendre polynomials.

\subsubsection{Herglotz's generating function}
Herglotz's generating function is a way to generate the spherical harmonics through a differentiation process: A plane wave
\begin{equation}
\exp(a_i x^i) = 
\sqrt{4\pi}\sum_{\ell = 0}^\infty\sum_{m=-\ell}^{+\ell}
\frac{\lambda^m r^\ell Y_{\ell m}(\theta,\varphi)}{\sqrt{(2\ell+1)(\ell+m)!(\ell-m)!}}
\end{equation}
is explained as a sum over spherical harmonics, where the wave has a wave vector $a_i$ and the coordinate $x^i$,
\begin{equation}
a_i = \left(\begin{array}{ccc} 
\frac{1}{2\lambda}-\frac{\lambda}{2}, & -\frac{\ci}{2\lambda}-\frac{\ci\lambda}{2}, & 1\end{array}\right)
\quad\mathrm{and}\quad
x^i = r\left(\begin{array}{c}\sin\theta\cos\varphi\\ \sin\theta\sin\varphi\\ \cos\theta\end{array}\right).
\end{equation}

The wave vector is peculiar as it has a norm of zero. For simplification, we focus on the choice $\lambda = 1$, which leads to
\begin{equation}
\exp(a_i x^i) = 
\sqrt{4\pi}\sum_{\ell = 0}^\infty\sum_{m=-\ell}^{+\ell}
\frac{r^\ell Y_{\ell m}(\theta,\varphi)}{\sqrt{(2\ell+1)(\ell+m)!(\ell-m)!}}
\end{equation}
with
\begin{equation}
a_i = \left(\begin{array}{ccc} 
0, & -\ci, & 1\end{array}\right)
\end{equation}
such that
\begin{equation}
a_ix^i = r\left(-\ci\sin\theta\sin\varphi + \cos\theta\right),
\end{equation}
yielding
\begin{multline}
\exp(a_ix^i) = \\
\exp(r(-\ci\sin\theta\sin\varphi + \cos\theta)) = \\
\exp(-\ci r\sin\theta\sin\varphi)\exp(r\cos\theta).
\end{multline}
Of particular interest is the value $\theta = \pi/2$, at which $\sin\theta = 1$ and $\cos\theta = 0$. Then, $\exp(a_ix^i) = \exp(-\ci r\sin\varphi) = \cos(r\sin\varphi) - \ci\sin(r\sin\varphi)$: Consequently, one would expect this functional dependence on the equator.

\begin{figure}[htb]
    \centering
    \begin{tabular}{c}
    \includegraphics[width=0.45\textwidth]{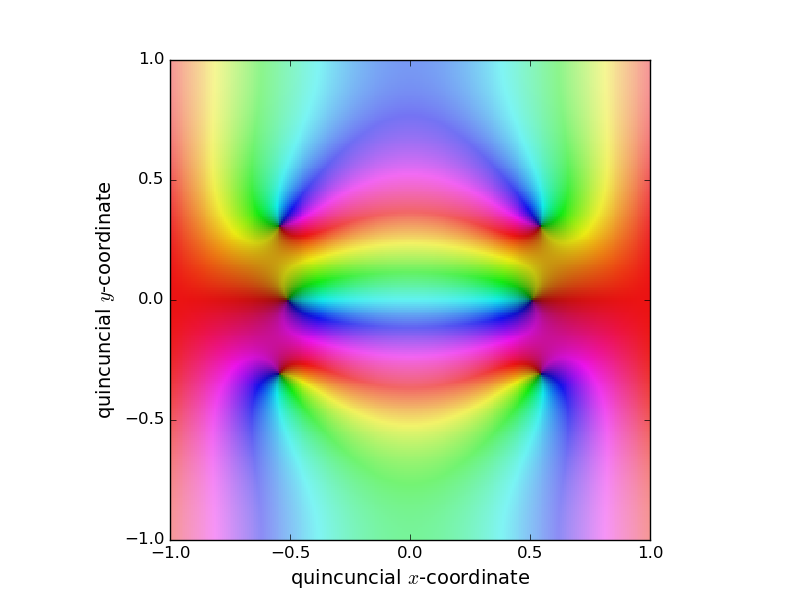} \\
    \includegraphics[width=0.45\textwidth]{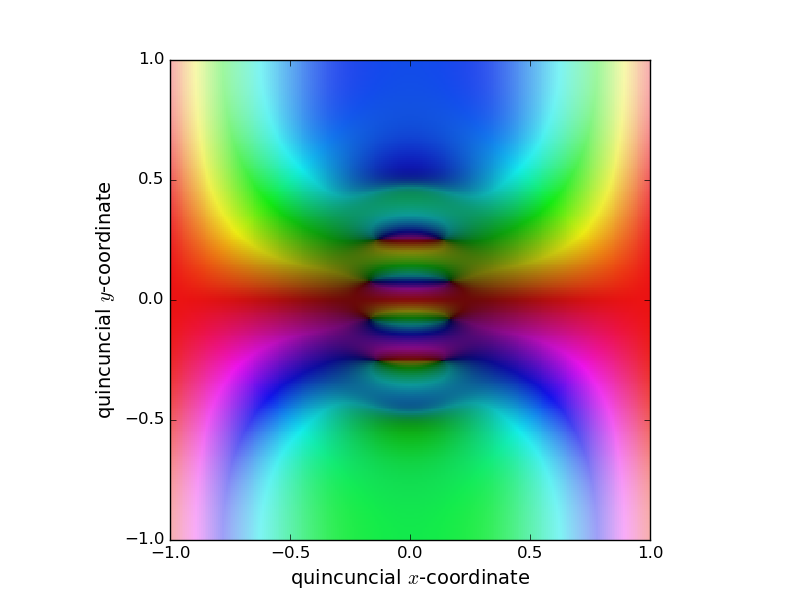} \\
    \includegraphics[width=0.45\textwidth]{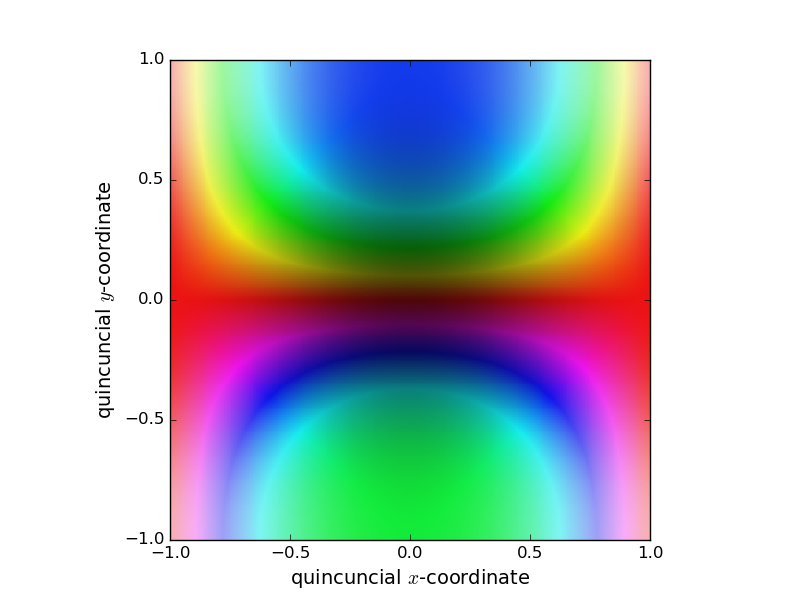}
    \end{tabular}
    \caption{Herglotz's generating function, shown as partial sums in $\ell$, and the target function $\exp(-\ci r\sin\theta\sin\varphi + r\cos\theta)$ for the particular choice $\lambda = 1$ (third panel), shown for $\ell_\mathrm{max} = 3$ (first panel), $\ell_\mathrm{max} = 10$ (second panel), all for the choice $r = 4$.}
    \label{fig:generating}
\end{figure}

The convergence of partial sums in $\ell$ towards the target $\exp(r(-\ci\sin\theta\sin\varphi + \cos\theta)) $ is shown in Fig.~\ref{fig:generating}, for different values of $\ell_\mathrm{max}$. The features around the North pole present at low $\ell_\mathrm{max}$ vanish rather quickly with increasing $\ell_\mathrm{max}$.

\subsubsection{Rayleigh-expansion}
The Rayleigh-expansion states that there is a superposition of spherical waves that gives rise to a plane wave,
\begin{equation}
\exp(\ci k_ir^i) = 
4\pi\sum_{\ell=0}^\infty\ci^\ell j_\ell(kr)\sum_{m=-\ell}^{+\ell}Y_{\ell m}(\hat{r})Y_{\ell m}^*(\hat{k}),
\label{eqn_rayleigh}
\end{equation}
which plays a role in the partial wave decomposition technique in scattering theory and in cosmology, particularly in applications such as $3d$ weak gravitational lensing \citep{heavens_3d_2003}.

While such a relation might appear surprising at first sight, it is readily explained by the fact that Euclidean space has both shifts and rotations as fundamental symmetries. $\hat{r}$ and $\hat{k}$ are shorthand notations of unit vectors associated with $r^i$ and $k_i$. Having the wave propagate into the $z$-direction with a unit wave vector $k_i = (0,0,1)$ gives $k_ir^i = r\cos\theta$, implying
\begin{equation}
4\pi\sum_{\ell=0}^\infty\ci^\ell j_\ell(kr)\sum_{m=-\ell}^{+\ell}Y_{\ell m}(\hat{r})Y_{\ell m}^*(\hat{k}) = 
\exp(\ci r\cos\theta),
\end{equation}
if $r^i$ has the norm $r_ir^i = r^2$.

\begin{figure}[htb]
    \centering
    \begin{tabular}{c}
    \includegraphics[width=0.45\textwidth]{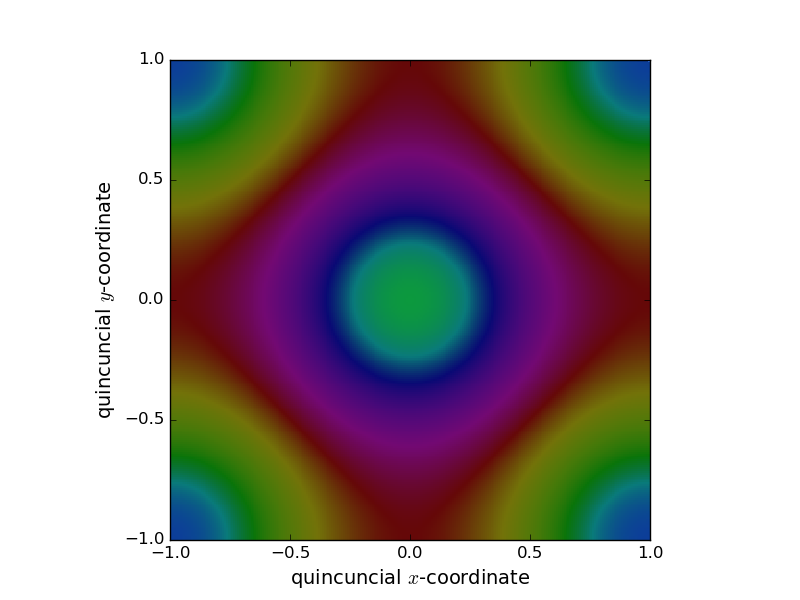} \\
    \includegraphics[width=0.45\textwidth]{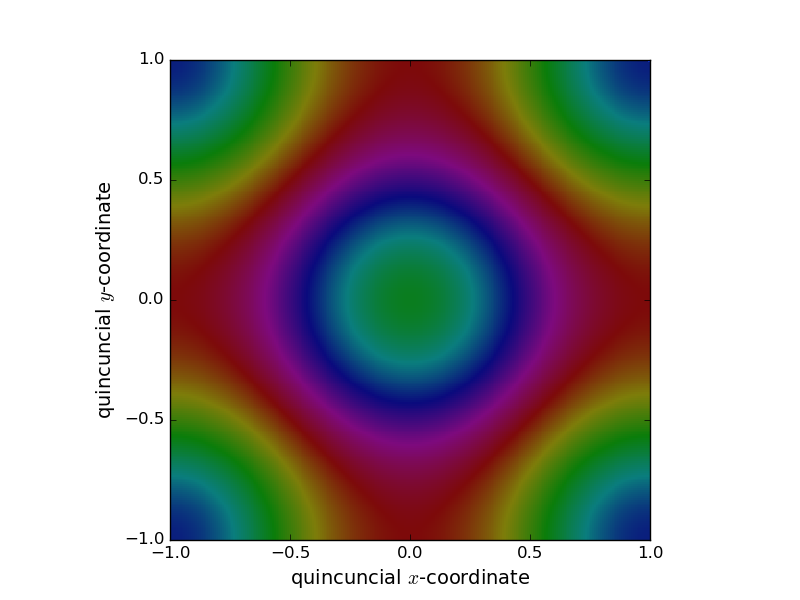} \\
    \includegraphics[width=0.45\textwidth]{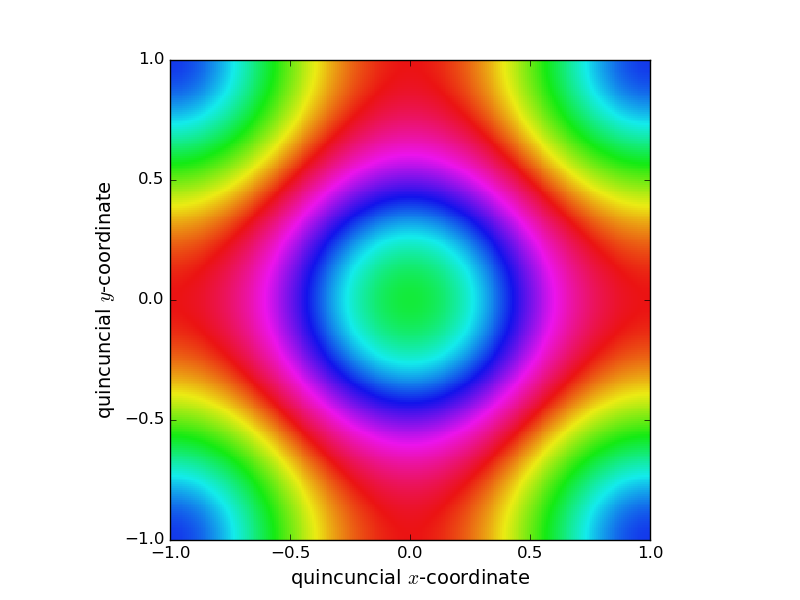}
    \end{tabular}
    \caption{Partial sums in $\ell$ for Rayleigh's expansion, which should successively give rise to $\exp(\ci r\cos\theta)$ (third panel), shown for $\ell_\mathrm{max} = 3$ (first panel), $\ell_\mathrm{max} = 10$ (second panel), all for the choice $r = 4$.}
    \label{fig:rayleigh}
\end{figure}

Fig.~\ref{fig:rayleigh} demonstrates that cumulative evaluations of eqn.~(\ref{eqn_rayleigh}) approximate the plane wave better with increasing $\ell_\mathrm{max}$. Convergence is surprisingly fast, and already at low $\ell_\mathrm{max}$ one approximates the final plane wave remarkably well.

\subsubsection{Wigner $3j$-symbols}
While the integral over two spherical harmonics is simply the orthonormality relation, the integral over a product of three spherical harmonics
\begin{multline}
\int_{4\pi}\dd\Omega\:Y_{\ell_1m_1}(\theta,\varphi)Y_{\ell_2m_2}(\theta,\varphi)Y_{\ell_3m_3}(\theta,\varphi) = \\
\sqrt{\frac{(2\ell_1+1)(2\ell_2+1)(2\ell_3+1)}{4\pi}}
\left(
\begin{array}{ccc}
\ell_1 & \ell_2 & \ell_3 \\ 0 & 0 & 0
\end{array}
\right)
\left(
\begin{array}{ccc}
\ell_1 & \ell_2 & \ell_3 \\ m_1 & m_2 & m_3
\end{array}
\right)
\label{eqn_wigner}
\end{multline}
gives rise to the Wigner-$3j$ symbols: There are closed forms for evaluation, but more importantly, there are straightforward rules under what circumstances the right hand side of eqn.~(\ref{eqn_wigner}) can at all be unequal to zero: The sum of all $m_i$ needs to vanish, $m_1+m_2+m_3 = 0$, the $\ell_i$ need to fulfil the triangle inequality, $|\ell_1 - \ell_2| \leq \ell_3\leq \ell_1+\ell_2$, and the sum of all $\ell_i$, $\ell_1+\ell_2 + \ell_3$, needs to be an even integer.

\begin{figure}[htb]
    \centering
    \begin{tabular}{c}
    \includegraphics[width=0.45\textwidth]{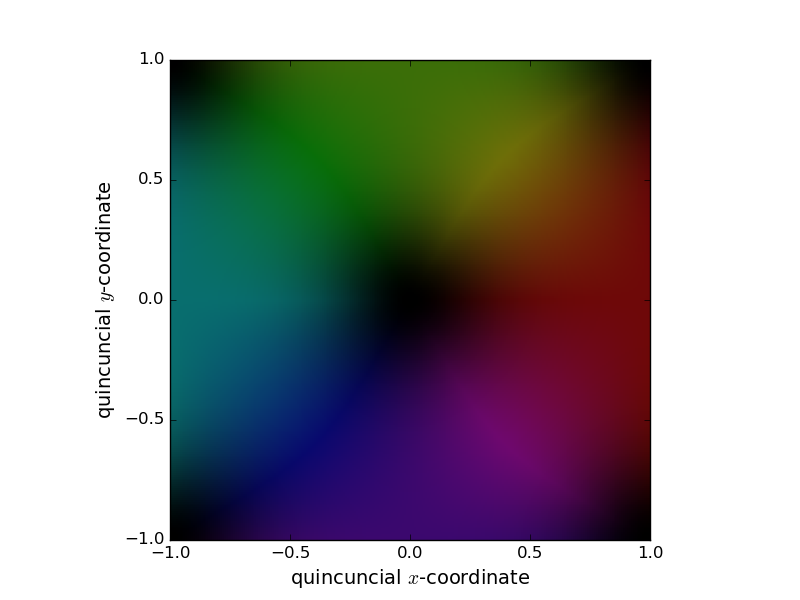} \\
    \includegraphics[width=0.45\textwidth]{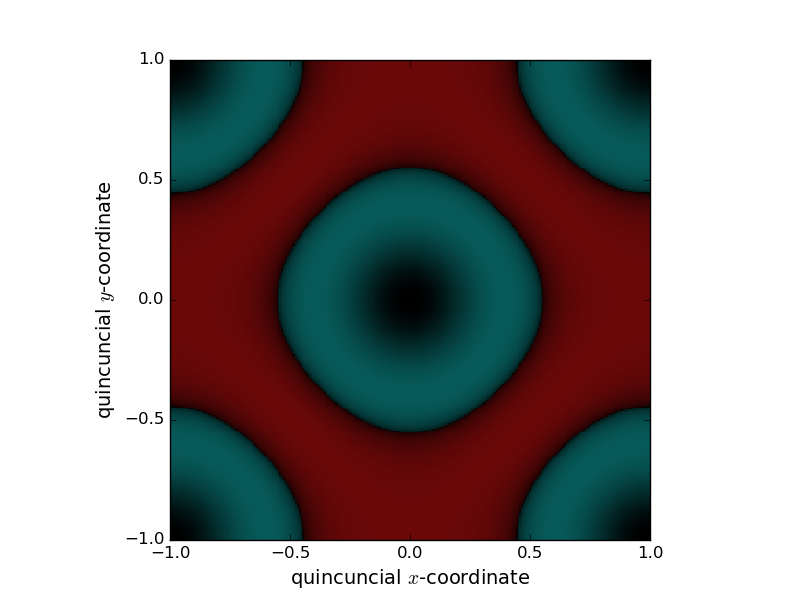} \\
    \includegraphics[width=0.45\textwidth]{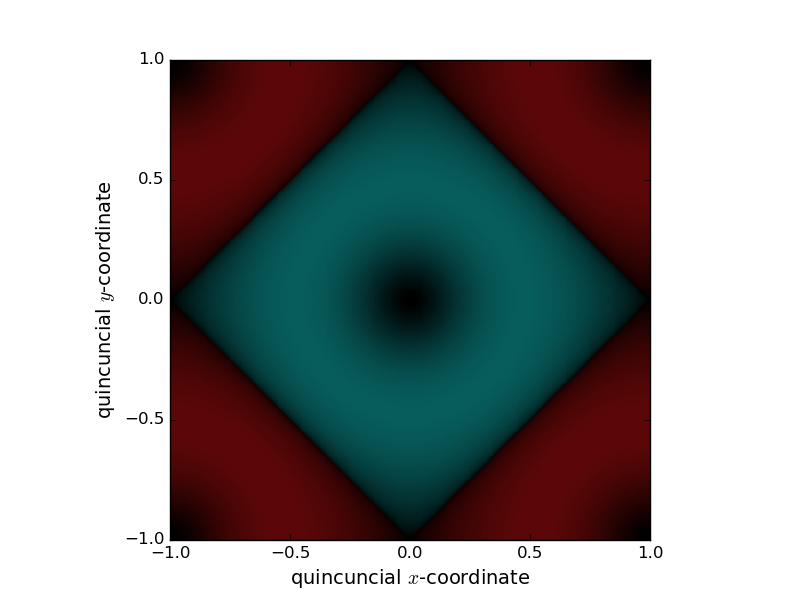}
    \end{tabular}
    \caption{Product of three spherical harmonics, where the choice $Y_{11}(\theta,\varphi)Y_{22}(\theta,\varphi)Y_{2,-2}(\theta,\varphi)$ in the first panel violates the condition $m_1+m_2+m_3 = 0$, with the choice $Y_{11}(\theta,\varphi)Y_{22}(\theta,\varphi)Y_{5,-3}(\theta,\varphi)$ in the second panel in contradiction with the triangle inequality, and for the case $Y_{11}(\theta,\varphi)Y_{21}(\theta,\varphi)Y_{2,-2}(\theta,\varphi)$, where the sum $\ell_1+\ell_2+\ell_3$ is odd instead of even.}
    \label{fig:wigner}
\end{figure}

Fig.~\ref{fig:wigner} shows three cases where these conditions are individually violated, such that the integral in eqn.~(\ref{eqn_wigner}) vanishes. The first case, where $m_1 + m_2 + m_3 = 1$ instead of zero, gives rise to a complex valued function $\propto\exp(\ci\varphi)$ with a single cycle through the colours. Consequently, one finds a cancellation of all contributions to the integral at $\varphi$ and $\varphi + \pi$. The second case deals with a violation of the triangle inequality, as $\ell_3 = 5$ does not fulfil $|\ell_1 - \ell_2| \leq \ell_3\leq \ell_1+\ell_2$ with $\ell_1 = 1$ and $\ell_2 = 2$. The sum of all $m$ is zero, such that the integrand is real-valued, and with similar areas of opposite signs one can guess the integration to vanish. A clearer case is the third, where the sum $\ell_1 + \ell_2 + \ell_3 = 5$ is odd instead of even, and the visualisation shows a Northern hemisphere with opposite sign to the Southern hemisphere, again leading to a zero result.

\subsubsection{Recovery of Fourier-transforms in the small-angle limit}
On small scales, the spherical harmonics $Y_{\ell m}(\theta,\varphi)$ should approximate plane waves at high $\ell$, as can be seen by this argument: The actual defining differential equation
\begin{equation}
\Delta Y_{\ell m}(\theta,\varphi) = -\ell(\ell+1)Y_{\ell m}(\theta,\varphi)
\end{equation}
transitions to 
\begin{equation}
\Delta Y_{\ell m}(\theta,\varphi) = -\ell^2Y_{\ell m}(\theta,\varphi),
\quad\mathrm{for}\quad\ell\gg 1
\end{equation}
in recovery of the Helmholtz differential equation which has plane waves as solutions, with wave number $\ell$. Intuitively, this means that on small scales $\theta\ll\pi$, that are probed when $\ell\gg 1/\pi$, the curvature of the sphere can be neglected and the sphere can be locally approximated by a tangential plane. On that plane, the harmonic decomposition eqn.~(\ref{ylm_analysis}) falls back onto a Fourier-transform. While plane waves are a part of the $Y_{\ell m}(\theta,\varphi)$ in the $\varphi$-direction for any index, the Legendre polynomials are well approximated by sinusoidal waves for $\ell\gg 1$, as their amplitudes become constant and their zeros equidistant.

\begin{figure}[htb]
    \centering
    \begin{tabular}{cc}
    \includegraphics[width=0.45\textwidth]{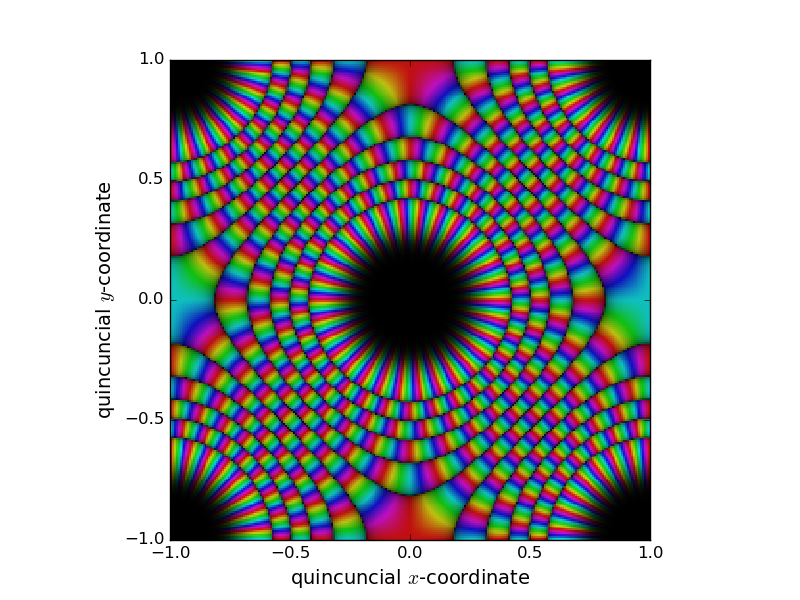}
    \end{tabular}
    \caption{Spherical harmonics $Y_{\ell m}(\theta,\varphi)$ for very high $\ell = 40$, where the small angle approximation is applicable, for a choice of $m = 30$ corresponding to the tesseral type. At the centres of the sides of the map one clearly sees the distortion induced by the variability of the conformal factor.}
    \label{fig:fourier}
\end{figure}

This is shown in Fig.~\ref{fig:fourier} for the combination $\ell = 40$ and $m = 30$. Typical wave lengths amount to a few degrees, and one sees the checkered pattern generated by this tesseral spherical harmonic. At the same time, one clearly sees how Peirce's projection has relatively small distortions along the diagonals, but relatively strong distortions at the centres of the sides, where the checker pattern appears enlarged.

\subsubsection{Unitary time evolution in quantum mechanics}
Time evolution of wave functions in quantum mechanics is accomplished by a unitary operator $U(t,t_0)$,
\begin{equation}
U(t,t_0) = \exp\left(-\frac{2\pi\ci H(t-t_0)}{h}\right)
\end{equation}
acting on the wave function
\begin{equation}
\psi(t) = U(t,t_0) \psi(t_0),
\end{equation}
with Planck's constant $h$ and the Hamilton-operator $H$. If the system is in an energy eigenstate, the phase of the wave function evolves linearly with time by virtue of $U(t,t_0)$. Unitary time-evolution applied to \href{https://heibox.uni-heidelberg.de/f/d4f73798cf684b87b3a8/}{$Y_{77}(\theta,\varphi)$} shows a smooth flow of the colours in azimuth. The (stationary) probability current associated with this phase evolution gives rise to the magnetic moment, which could be measured in e.g. a hydrogen atom with the quantum number $m = 7$.

\section{Summary}
Peirce's quincuncial projection allows visualisations of functions defined on the surface of the sphere, in particular of the spherical harmonics $Y_{\ell m}(\theta,\varphi)$, which find widespread use in physics, geophysics and astronomy. In particular the complex structure of the $Y_{\ell m}(\theta,\varphi)$ and their symmetries can be illustrated, in conjunction with their fundamental properties that make them attractive for usage throughout physics: $(i)$ hermiticity, $(ii)$ parity, $(iii)$ orthonormality, $(iv)$ completeness, $(v)$ Herglotz's generating function, $(vi)$ the Rayleigh expansion of plane waves into spherical waves, $(vii)$ the Wigner $3j$-symbols and finally $(viii)$ their  relation to Fourier-transforms. There are advantages of representing the spherical harmonics in a flat map projection in some cases, and it makes their functional form accessible to students who find it difficult to visualise $3d$-structures.

\section{Acknowledgements}
I thank Chamberlain Fong for encouragement to write this article. The {\em python}-scripts used for generating the plots are available on request from the author. I acknowledge the implementation of the quincuncial projection by C. Fong \citep{DBLP:journals/corr/abs-1011-3189, fong2024analyticalmethodssquaringdisc}, which I translated from {\em matlab} to {\em python}. Visualisations of the phase of complex function by colour are difficult to process by readers with a divergent colour perception: I am working on plotting scripts in {\em python} for complex functions that use a perceptually uniform colormap. Numerical implementations for Jacobi's elliptical functions were the ones in {\em SciPy}, and for the spherical harmonics I used the {\em python} package {\em Spheroidal}. The data set of the spherical harmonics expansion coefficients of the Earth was kindly provided by Matthias Bartelmann, and I thank Tristan Bereau for comments on the topological map. I am grateful to Theodora Lazarevic and Justus Zimmermann for spotting mistakes.

\bibliographystyle{spbasic}
\bibliography{references}

\begin{thebibliography}{32}
\providecommand{\natexlab}[1]{#1}
\providecommand{\url}[1]{{#1}}
\providecommand{\urlprefix}{URL }
\expandafter\ifx\csname urlstyle\endcsname\relax
  \providecommand{\doi}[1]{DOI~\discretionary{}{}{}#1}\else
  \providecommand{\doi}{DOI~\discretionary{}{}{}\begingroup
  \urlstyle{rm}\Url}\fi
\providecommand{\eprint}[2][]{\url{#2}}

\bibitem[{Adamek et~al(2015)Adamek, Durrer, and Tansella}]{adamek_lensing_2015}
Adamek J, Durrer R, Tansella V (2015) Lensing signals from spin-2
  perturbations. arxiv/151001566
  \urlprefix\url{http://arxiv.org/abs/1510.01566}, \eprint{1510.01566}

\bibitem[{Beck et~al(2018)Beck, Fabbian, and Errard}]{beck_lensing_2018}
Beck D, Fabbian G, Errard J (2018) Lensing reconstruction in post-born cosmic
  microwave background weak lensing. arxiv/180601216
  \urlprefix\url{http://arxiv.org/abs/1806.01216}, \eprint{1806.01216}

\bibitem[{Christoffel(1867)}]{christoffel_1867_2358602}
Christoffel EB (1867) Sul problema delle temperature stazionarie e la
  rappresentazione di una data superficie. \doi{10.1007/bf02419161},
  \urlprefix\url{https://doi.org/10.1007/bf02419161}

\bibitem[{Dray(1985)}]{10.1063/1.526533}
Dray T (1985) The relationship between monopole harmonics and spin‐weighted
  spherical harmonics. Journal of Mathematical Physics 26(5):1030--1033,
  \doi{10.1063/1.526533}, \urlprefix\url{https://doi.org/10.1063/1.526533},
  \eprint{https://pubs.aip.org/aip/jmp/article-pdf/26/5/1030/19220480/1030\_1\_online.pdf}

\bibitem[{Driscoll and Trefethen(2002)}]{Driscoll_Trefethen_2002}
Driscoll TA, Trefethen LN (2002) Schwarz-Christoffel Mapping. Cambridge
  Monographs on Applied and Computational Mathematics, Cambridge University
  Press

\bibitem[{Fern{\'a}ndez-Cobos et~al(2016)Fern{\'a}ndez-Cobos, Marcos-Caballero,
  Vielva, Mart{\'\i}nez-Gonz{\'a}lez, and
  Barreiro}]{fernandez-cobos_exploring_2016}
Fern{\'a}ndez-Cobos R, Marcos-Caballero A, Vielva P, Mart{\'\i}nez-Gonz{\'a}lez
  E, Barreiro RB (2016) Exploring 2-spin internal linear combinations for the
  recovery of the {CMB} polarization. arxiv/160101515
  \urlprefix\url{http://arxiv.org/abs/1601.01515}, \eprint{1601.01515}

\bibitem[{Fong(2010)}]{DBLP:journals/corr/abs-1011-3189}
Fong C (2010) Warping peirce quincuncial panoramas. CoRR abs/1011.3189,
  \urlprefix\url{http://arxiv.org/abs/1011.3189}, \eprint{1011.3189}

\bibitem[{Fong(2024)}]{fong2024analyticalmethodssquaringdisc}
Fong C (2024) Analytical methods for squaring the disc.
  \urlprefix\url{https://arxiv.org/abs/1509.06344}, \eprint{1509.06344}

\bibitem[{Gluscevic et~al(2009)Gluscevic, Kamionkowski, and
  Cooray}]{gluscevic_-rotation_2009}
Gluscevic V, Kamionkowski M, Cooray A (2009) De-rotation of the cosmic
  microwave background polarization: Full-sky formalism. Physical Review D
  80(2), \doi{10.1103/PhysRevD.80.023510},
  \urlprefix\url{http://arxiv.org/abs/0905.1687}, \eprint{0905.1687}

\bibitem[{Goldberg et~al(1967)Goldberg, Macfarlane, Newman, Rohrlich, and
  Sudarshan}]{goldberg_spins_1967}
Goldberg JN, Macfarlane AJ, Newman ET, Rohrlich F, Sudarshan ECG (1967)
  Spin‐s spherical harmonics and $\eth$. Journal of Mathematical Physics
  8(11):2155--2161, \doi{10.1063/1.1705135},
  \urlprefix\url{http://aip.scitation.org/doi/abs/10.1063/1.1705135}

\bibitem[{Gorski et~al(1999)Gorski, Wandelt, Hansen, Hivon, and
  Banday}]{gorski1999healpixprimer}
Gorski KM, Wandelt BD, Hansen FK, Hivon E, Banday AJ (1999) The healpix primer.
  \urlprefix\url{https://arxiv.org/abs/astro-ph/9905275},
  \eprint{astro-ph/9905275}

\bibitem[{Gorski et~al(2005)Gorski, Hivon, Banday, Wandelt, Hansen, Reinecke,
  and Bartelmann}]{Gorski_2005}
Gorski KM, Hivon E, Banday AJ, Wandelt BD, Hansen FK, Reinecke M, Bartelmann M
  (2005) Healpix: A framework for high‐resolution discretization and fast
  analysis of data distributed on the sphere. The Astrophysical Journal
  622(2):759--771, \doi{10.1086/427976},
  \urlprefix\url{http://dx.doi.org/10.1086/427976}

\bibitem[{Hall and Taylor(2014)}]{hall_intrinsic_2014}
Hall A, Taylor A (2014) Intrinsic alignments in the cross-correlation of cosmic
  shear and cosmic microwave background weak lensing. MNRAS 443:L119--L123,
  \doi{10.1093/mnrasl/slu094}

\bibitem[{Hanson et~al(2010)Hanson, Challinor, and Lewis}]{hanson_weak_2010}
Hanson D, Challinor A, Lewis A (2010) Weak lensing of the {CMB}. General
  Relativity and Gravitation 42(9):2197--2218, \doi{10.1007/s10714-010-1036-y},
  \urlprefix\url{http://link.springer.com/10.1007/s10714-010-1036-y}

\bibitem[{Heavens(2003)}]{heavens_3d_2003}
Heavens A (2003) {3D} weak lensing. {\textbackslash}mnras 343:1327--1334,
  \doi{10.1046/j.1365-8711.2003.06780.x}

\bibitem[{Hirata and Seljak(2003)}]{hirata_reconstruction_2003}
Hirata CM, Seljak U (2003) Reconstruction of lensing from the cosmic microwave
  background polarization. Physical Review D 68(8):083,002--+

\bibitem[{Hudson(1978)}]{10.1093/gji/52.2.366}
Hudson JA (1978) Physics of the earth f. d. stacey, john wiley \&amp; sons,
  1977 414 pages. Geophysical Journal International 52(2):366--366,
  \doi{10.1093/gji/52.2.366},
  \urlprefix\url{https://doi.org/10.1093/gji/52.2.366},
  \eprint{https://academic.oup.com/gji/article-pdf/52/2/366/1649642/52-2-366.pdf}

\bibitem[{Huffenberger and Wandelt(2010)}]{Huffenberger_2010}
Huffenberger KM, Wandelt BD (2010) Fast and exact spin-s spherical harmonic
  transforms. The Astrophysical Journal Supplement Series 189(2):255,
  \doi{10.1088/0067-0049/189/2/255},
  \urlprefix\url{https://dx.doi.org/10.1088/0067-0049/189/2/255}

\bibitem[{Hughes(2000)}]{PhysRevD.61.084004}
Hughes SA (2000) Evolution of circular, nonequatorial orbits of kerr black
  holes due to gravitational-wave emission. Phys Rev D 61:084,004,
  \doi{10.1103/PhysRevD.61.084004},
  \urlprefix\url{https://link.aps.org/doi/10.1103/PhysRevD.61.084004}

\bibitem[{Kamionkowski and Kovetz(2015)}]{kamionkowski_quest_2015}
Kamionkowski M, Kovetz ED (2015) The quest for b modes from inflationary
  gravitational waves. arxiv/151006042
  \urlprefix\url{http://arxiv.org/abs/1510.06042}, \eprint{1510.06042}

\bibitem[{Leistedt et~al(2015)Leistedt, {McEwen}, Kitching, and
  Peiris}]{leistedt_3d_2015}
Leistedt B, {McEwen} JD, Kitching TD, Peiris HV (2015) 3d weak lensing with
  spin wavelets on the ball. arxiv/150906750
  \urlprefix\url{http://arxiv.org/abs/1509.06750}, \eprint{1509.06750}

\bibitem[{Lewis and Challinor(2007)}]{lewis_21cm_2007}
Lewis A, Challinor A (2007) The 21cm angular-power spectrum from the dark ages.
  Physical Review D 76(8), \doi{10.1103/PhysRevD.76.083005},
  \urlprefix\url{http://arxiv.org/abs/astro-ph/0702600},
  \eprint{astro-ph/0702600}

\bibitem[{Maino et~al(1999)Maino, Burigana, Maltoni, Wandelt,
  G\{{\textbackslash}textbackslash\}'~orski, Malaspina, Bersanelli, Mandolesi,
  Banday, and Hivon}]{maino_planck-lfi_1999}
Maino D, Burigana C, Maltoni M, Wandelt BD,
  G\{{\textbackslash}textbackslash\}'~orski KM, Malaspina M, Bersanelli M,
  Mandolesi N, Banday AJ, Hivon E (1999) The planck-{LFI} instrument: Analysis
  of the 1/f noise and implications for the scanning strategy. AAPS
  140:383--391

\bibitem[{Merkel and Schaefer(2011)}]{merkel_gravitational_2011}
Merkel P, Schaefer BM (2011) Gravitational lensing of the cosmic microwave
  background by nonlinear structures. Monthly Notices of the Royal Astronomical
  Society 411(2):1067--1076, \doi{10.1111/j.1365-2966.2010.17739.x},
  \urlprefix\url{http://arxiv.org/abs/1007.1408}, \eprint{1007.1408}

\bibitem[{Nishizawa et~al(2008)Nishizawa, Komatsu, Yoshida, Takahashi, and
  Sugiyama}]{nishizawa_cosmic_2008}
Nishizawa AJ, Komatsu E, Yoshida N, Takahashi R, Sugiyama N (2008) Cosmic
  microwave background--weak lensing correlation: Analytical and numerical
  study of non-linearity and implications for dark energy. The Astrophysical
  Journal 676(2):L93--L96, \doi{10.1086/587741},
  \urlprefix\url{http://arxiv.org/abs/0711.1696}, \eprint{0711.1696}

\bibitem[{Okamoto and Hu(2003)}]{okamoto_cosmic_2003}
Okamoto T, Hu W (2003) Cosmic microwave background lensing reconstruction on
  the full sky. Physical Review D 67(8):083,002--+,
  \doi{10.1103/PhysRevD.67.083002}

\bibitem[{Olver et~al(2025)Olver, {Olde Daalhuis}, Lozier, Schneider, Boisvert,
  Clark, Miller, Saunders, Cohl, and McClain}]{NIST:DLMF}
Olver FWJ, {Olde Daalhuis} AB, Lozier DW, Schneider BI, Boisvert RF, Clark CW,
  Miller BR, Saunders BV, Cohl HS, McClain MA (2025) Nist digital library of
  mathematical functions. \url{https://dlmf.nist.gov/}, Release 1.2.4 of
  2025-03-15, \urlprefix\url{https://dlmf.nist.gov/}

\bibitem[{Peirce(1879)}]{c14fbc06-496f-3cbc-8158-f8078819cb05}
Peirce CS (1879) A quincuncial projection of the sphere. American Journal of
  Mathematics 2(4):394--396,
  \urlprefix\url{http://www.jstor.org/stable/2369491}

\bibitem[{Schwarz(1869)}]{Schwarz+1869+105+120}
Schwarz H (1869) {\"U}ber einige abbildungsaufgaben. Journal f{\"u}r die reine
  und angewandte Mathematik 1869(70):105--120,
  \doi{doi:10.1515/crll.1869.70.105},
  \urlprefix\url{https://doi.org/10.1515/crll.1869.70.105}

\bibitem[{Seljak(1996)}]{seljak_gravitational_1996}
Seljak U (1996) Gravitational lensing effect on cosmic microwave background
  anisotropies: A power spectrum approach. ApJ 463:1, \doi{10.1086/177218}

\bibitem[{Solanilla et~al(2016)Solanilla, Oostra, and
  Yanez}]{solanilla_mapping}
Solanilla L, Oostra A, Yanez J (2016) Peirce quincuncial projection, Revista
  Integracion, vol~34. Escuela de Matematicas

\bibitem[{Zaldarriaga and Seljak(1998)}]{zaldarriaga_gravitational_1998}
Zaldarriaga M, Seljak U (1998) Gravitational lensing effect on cosmic microwave
  background polarization. Physical Review D 58(2):023,003--+,
  \doi{10.1103/PhysRevD.58.023003}

\end{thebibliography}

\appendix
\section{Projection formulas for Peirce's quincuncial projection}\label{sect_appendix}
The explicit relation for projecting a point on the square domain $(x,y)\in [-1,+1]\times[-1,+1]$ to coordinates $x,y$ the stereographic plane is given by
\begin{equation}
u = K(m)(x-1)
\quad\mathrm{and}\quad
w = K(m) y
\end{equation}
with the complete elliptic integral
\begin{equation}
K(m) = \int_0^{\pi/2}\dd\vartheta\:\frac{1}{\sqrt{1-m\sin^2\vartheta}}
\end{equation}
where we need the specific value $m = 1/2$. Furthermore, the Jacobi elliptic functions are implicitly defined through the relations
\begin{equation}
\mathrm{sn}(u(\varphi),m) = \sin\varphi,
\quad
\mathrm{cn}(u(\varphi),m) = \cos\varphi,
\quad
\mathrm{dn}(u(\varphi),m) = \sqrt{1-m^2\sin^2\varphi}
\end{equation}
with the elliptic integral
\begin{equation}
u(\varphi) = \int_0^\varphi\dd\vartheta\:\frac{1}{\sqrt{1-m^2\sin^2\vartheta}}
\end{equation}

Then, the abbreviations
\begin{equation}
s = \mathrm{sn}(u,m),\quad
c = \mathrm{cn}(u,m),\quad
d = \mathrm{dn}(u,m)
\end{equation}
as well as
\begin{equation}
s_1 = \mathrm{sn}(w,1-m),\quad
c_1 = \mathrm{cn}(w,1-m),\quad
d_1 = \mathrm{dn}(w,1-m)
\end{equation}
are used to compute
\begin{equation}
\delta = c_1^2 + m (s s_1)^2
\end{equation}
from which the stereographic coordinates follow as 
\begin{equation}
x = \frac{c c_1}{\delta}
\quad\mathrm{and}\quad
y = -\frac{s s_1 \times d d_1}{\delta}.
\end{equation}
		
Inverse stereographic projection then determines latitude $\lambda$ and longitude $\beta$ through 
\begin{equation}
	\lambda = 2\arctan\sqrt{x^2+ y^2} - \pi / 2
\quad\mathrm{and}\quad
	\beta = \arctan(y/x),
\end{equation}
and finally, spherical coordinates with the polar angle $\theta = \pi/2 - \lambda$ and the azimuthal angle $\varphi = \beta + \pi$ follow directly.

\end{document}